# Using phase data from MR temperature imaging to visualize anatomy during MRI-guided focused ultrasound neurosurgery

Nathan McDannold, P. Jason White, and G. Rees Cosgrove

*Abstract*— Neurosurgery targets in the thalamus can be challenging to identify during transcranial MRI-guided focused ultrasound (MRgFUS) thermal ablation due to poor image quality. They also neighbor structures that can result in side effects if damaged. Here we demonstrate that the phase data obtained during MRgFUS for MR temperature imaging (MRTI) contains anatomic information that could be useful in guiding this procedure. This approach was evaluated in 68 thalamotomies for essential tremor (ET). We found that we could readily visualize the red nucleus and subthalamic nucleus, and those nuclei were consistently aligned with the sonication target coordinates. We also could consistently visualize the internal capsule, which needs to be protected from thermal damage to prevent side effects. Preliminary results from four pallidotomies in Parkinson's disease patients suggest that this approach might also be useful in visualizing the optic tract in addition to the internal capsule. Overall, this approach can visualize anatomic landmarks that may be useful to refine atlas-based targeting for MRgFUS. Since the same data is used for MRTI and anatomic visualization, there are no errors induced by registration to previously obtained planning images or image distortion, and no additional time is needed.

## I. INTRODUCTION

THALAMOTOMY with transcranial MRI-guided focused ultrasound (MRgFUS) thermal ablation is a noninvasive surgical approach that is approved for the treatment of essential tremor (ET) and tremor-dominant Parkinson's disease (PD) [2, 3]. MRgFUS thalamotomy for neuropathic pain, pallidotomy for PD, and capsulotomy for obsessive compulsive disorder, and other indications are being investigated [5-7]. Effectively treating these disorders without invasive surgery or ionizing radiation is a desirable option for many patients compared to deep brain stimulation, radiofrequency lesioning, or gamma knife radiosurgery.

Currently, the target for thalamotomy is selected based upon human stereotactic atlases and by measurements relative to anatomic structures such as the anterior commissure (AC), the posterior commissure (PC), and wall of the third ventricle. The target is typically selected on MRI scans acquired with the body coil with the patient in the MRgFUS device. The device itself and the water used to cool the head and provide acoustic coupling interfere with imaging leading to poor image quality. As a result, relevant anatomic structures in and around the thalamus are not well visualized.

Knowing the locations of these structures is also important to avoid side effects. The MRgFUS focal lesion is often oblique [8] and heating can extend into neighboring structures such as the internal capsule (IC). Improved anatomic imaging could also be useful in refining the sonication target, as atlas-based measurements alone do not take into account individual variability. Atlas-based planning can be enhanced using previously obtained high resolution imaging that is registered to the treatment planning images. Such imaging can allow for better visualization of relevant anatomy, but targeting based on it is only as good as the registration accuracy and can be susceptible to image distortion. While sub-millimeter registration error can be achieved under ideal conditions [9], the accuracy of the registration with the poor-quality images currently used with MRgFUS is not known. Furthermore, the requirement of a pre-treatment MRI for planning purposes is an added expense. Methods to refine both atlas-based targeting and direct targeting based on pre-treatment imaging would be beneficial.

The accuracy of the targeting and the accumulated thermal dose [10] during MRgFUS are monitored using MR temperature imaging (MRTI). This imaging is based on the temperature sensitivity of the proton resonant frequency of water [11]. Changes in resonant frequency are estimated using phase-difference images created with a gradient echo pulse sequence [12]. MRTI is obtained at multiple time points to track the heating during each sonication.

The phase of an MR image also contains anatomic information [13]. Structures such as the red nucleus (RN), substantia nigra (SN), and the subthalamic nucleus (STN),

This work was supported by NIH grant R01EB205205.

N. McDannold is with the Radiology Department at The Brigham and Women's Hospital and Harvard Medical School, Boston, MA 02115 USA (e-mail: njm@bwh.harvard.edu).

P.J. White is with the Radiology Department at The Brigham and Women's Hospital and Harvard Medical School, Boston, MA 02115 USA

(e-mail: white@ bwh.harvard.edu).

R. Cosgrove is with the Neurosurgery Department at The Brigham and Women's Hospital and Harvard Medical School, Boston, MA 02115 USA (e-mail: rcosgrove2@bwh.harvard.edu)

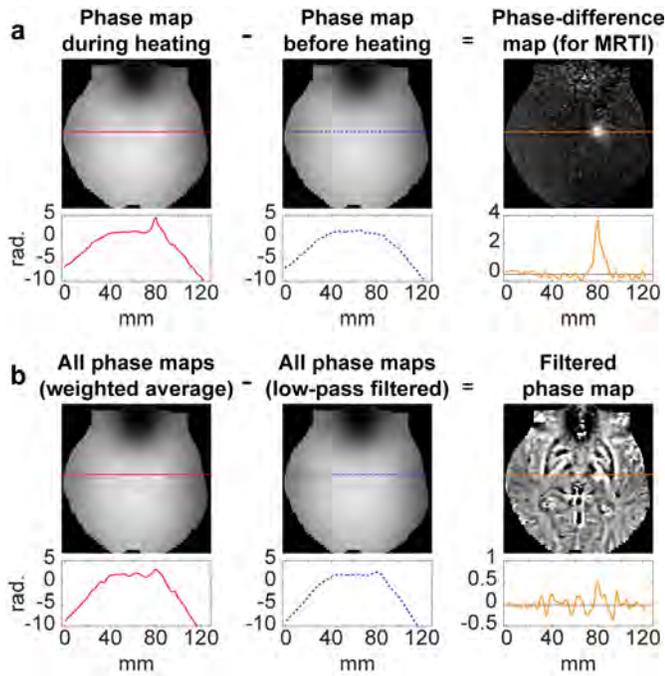

Fig. 1 Processing phase data obtained during sonication in two ways. (a) For MRTI, the phase maps acquired before heating are subtracted from those obtained at multiple time points during and after sonication. (b) To create the filtered phase maps, we first averaged all the phase maps acquired at all time points to increase the SNR. Anatomic information is faintly visible in the unwrapped phase map. A low pass filter of this data is created and then subtracted, and the resulting filtered phase map shows structural information with high contrast in the striatum and thalamus. Phase changes in the focal regions due to focal heating were not fully removed. Note that the heat-induced phase change in this example (focal temperature rise: 29.4°C) was larger than those associated with anatomic differences in magnetic susceptibility. Axial imaging is shown.

contain higher concentrations of iron [14]. This iron in the tissue consists largely of ferritin and hemosiderin and is paramagnetic. The magnetic field thus becomes slightly increased in these structures, which leads to a more phase accumulation between excitation and readout of the MRI data. The phospholipids in heavily myelinated white matter are diamagnetic, leading to decreased phase accumulation. These anatomic structures can be readily visualized by unwrapping the phase maps and removing contributions caused by inhomogeneities in the static magnetic field. These inhomogeneities have low spatial frequencies and can be removed with a high-pass filter [15, 16].

For tremor, the lesion is placed in the posteroventral portion of the ventralis intermedius (VIM) nucleus in the thalamus. This target is approximately aligned anteroposteriorly with the RN [1]. The VIM is adjacent to the internal capsule (IC) laterally and the zona incerta (ZI) inferiorly. The STN lies just below the ZI. Visualization of the RN, IC, and STN could thus be useful as landmarks to improve atlas-based targeting of the VIM and avoid damage to the IC.

In this work we show that the phase images acquired during sonication that are used for temperature mapping can reliably visualize these structures during thalamotomy for tremor. We retrospectively evaluated the imaging acquired during and after 68 MRgFUS thalamotomies and compared the average locations of these structures to the targets, the MRI-derived thermal dose, and the lesion at 24 h. We show that the locations of the nuclei in the phase maps are at locations expected from a stereotactic atlas, suggesting that these images could be useful to refine atlas-based targeting and avoiding damage to the IC. We also show preliminary results from four MRgFUS pallidotomies, where in addition to the IC, the optic tract ventral to the sonication target was detected.

## II. METHODS

### A. MRgFUS

We analyzed data from 68 MRgFUS procedures for ET and four for PD. For ET thalamotomy, the sonication target in the VIM nucleus. The nominal location of the target was 14 mm lateral to the midline, 25% of the ACPC distance anterior to the PC, and 1 mm superior to the ACPC plane. The neurosurgeon adjusted these locations based on the width of the third ventricle, the cranial size, and other considerations. For PD pallidotomy, the target was the globus pallidus interna (GPi), with a nominal location of 20 mm lateral to the midline, 3-4 mm anterior to the midpoint between AC and PC, and 3-4 mm inferior to the ACPC plane. The neurosurgeon adjusted those targets based on pre-treatment imaging and patient anatomy. The targeting is refined during treatment based on symptom improvement and reported side effects.

The treatments were performed using the ExAblate Neuro MRgFUS system (InSightec, Haifa, Israel). This system uses a hemispherical (diameter: 30 cm), 1024-element phased array transducer that operates at 660 kHz. The transducer was integrated into an MRI table for a 3T MRI (GE750W, GE Healthcare). The transducer was steered to within ±1 mm of the target selected by the neurosurgeon with a manual positioning system. Additional steering was provided by the phased array. The array was also used to correct for aberrations induced by the skull. This correction used information gleaned from CT scans obtained before treatment to predict and reverse skull-induced phase changes for each element [17]. The power for each element was set to normalize the intensity over the skull surface. The location and orientation of the transducer in the MRI space was obtained using MRI tracking coils.

Before treatment, the patient's head was completely shaved, and they were place in a stereotactic frame. A flexible waterproof membrane was stretched around the head. The patient was laid supine and the frame was attached securely to the MRI table. The membrane was attached to the open face of the MRgFUS transducer. The space between the scalp and transducer was filled with degassed and cooled water that was continuously circulated between sonications.

### B. MRI acquisition for temperature imaging

The targeting was verified using low energy (typically 1-3 kJ) sonications and MRTI. At least three verification sonications were applied to confirm targeting in each direction. Because of spatial distortions in the frequency-encoding direction, the verification only considered coordinates in the phase-encoding direction. After verification, the acoustic energy was increased to ablative levels in steps. Between each ablation sonication, the patient was withdrawn from the scanner and examined for symptom improvement and side effects. Based on these examinations, the sonication target was adjusted

## Axial

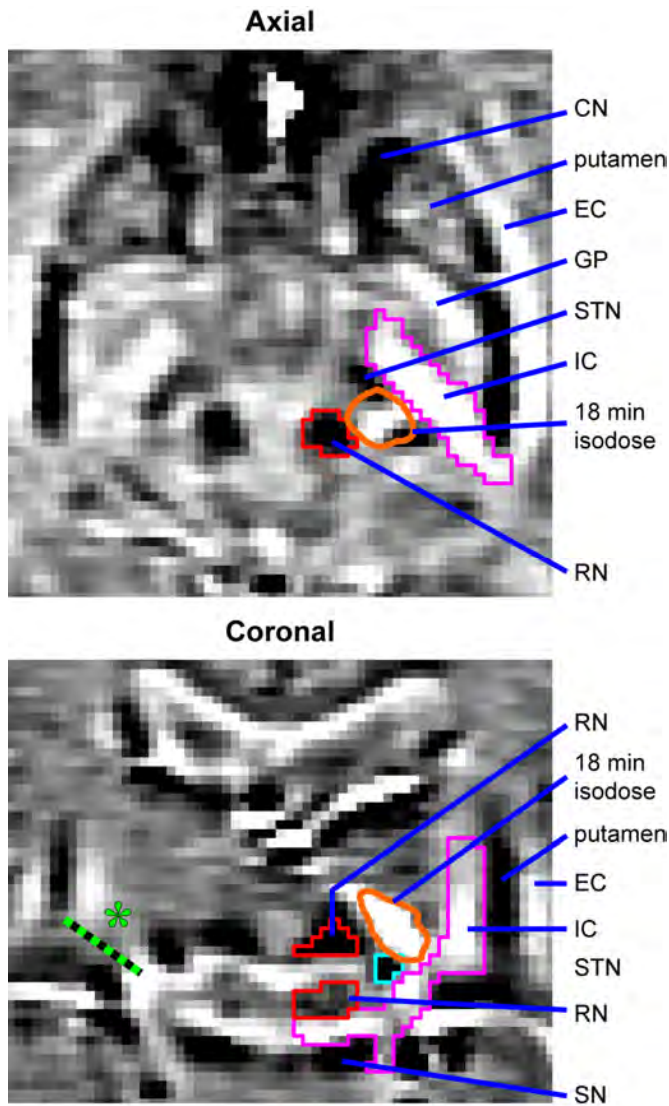

CN
putamen
EC
GP
STN
IC
18 min isodose
RN

## Coronal

RN
18 min isodose
putamen
EC
IC
STN
RN
SN

Fig. 2 Anatomic structures visible on the phase maps, along with thermal dose contours. The segmentations of the IC, RN, and STN are indicated. The lateroventral portion of the IC was often not visible in coronal imaging (*), perhaps due to its orientation at the "magic angle" [4] of 54.74° with respect to the magnetic field (dotted green line). The RN was separated into dorsal and ventral components. (CN: caudate nucleus; EC: external capsule; GP: globus pallidus; IC: internal capsule; RN: red nucleus; SN: substantia nigra; STN: subthalamic nucleus).

if necessary. Sonications were repeated until symptoms were substantially improved. Exposure levels were determined using MRTI, with an aim of achieving focal temperatures of ~51-60°C. Information about the sonication targeting and patient information is shown in Table S1.

Imaging for MRTI was obtained in a single plane using a spoiled gradient echo sequence. The orientation and frequency encoding direction of the MRTI was varied between sonications. For 17 treatments, a single echo readout was used (TR/TE: 27.8/12.9 ms; flip angle: 30°; receiver bandwidth ±5.7 kHz; field of view: 28 cm; slice thickness: 3 mm; matrix: 256×128). Otherwise the imaging was obtained using a multi-echo readout (TR: 28.0 ms; TE: 3.1/7.8/12.5/17.2/21.9 ms; flip angle: 30°; receiver bandwidth: ±35.71 kHz; field of view: 28 cm; slice thickness: 3 mm; matrix: 256×128). The low bandwidth of the single-echo imaging resulted in obvious distortions, often a millimeter or more, in the frequency encoding direction (Supplemental video 1).

Phase difference maps were generated with reference to imaging obtained before the start of sonication (Fig. 1a); imaging was repeated at multiple time points during heating and for five frames afterwards. Temporal changes in the phase difference maps unrelated to heating were removed by fitting non-heated areas to a smooth surface using a procedure outlined previously [18, 19]. With the multi-echo readout, we computed phase difference maps with respect to a pre-heating image for each TE and combined them using a weighted average with weights for the nth TE given by:

$$W_n = \left( TE_n \cdot e^{-TE_n/T2^*} \right)^2 \qquad (1)$$

We used 30 ms for the $T2^*$ relaxation time for brain.

The data were reconstructed as magnitude, real, and imaginary images in DICOM format. MRTI was generated using a temperature sensitivity -0.00909 ppm/°C [20], which was the value used by the MRgFUS software. The time series of MRTI was used to create maps of thermal dose [10].

The same data obtained for MRTI was used to create the filtered phase maps (Fig. 1b). First, we averaged all the complex images acquired during and after each sonication. After masking the images with segmentations of the brain, the phase was then unwrapped using a technique described elsewhere [21]. The segmentations were performed on the magnitude reconstructions of the MRTI. We removed the low spatial frequency components of the phase maps resulting from static magnetic field inhomogeneities, which we found using an averaging filter in the spatial domain with a kernel size of nine voxels (9.8 mm) [22]. With this filter, each voxel was averaged with its 80 neighbors in 9×9 voxel grid. This low pass filtered phase map was subtracted off, leaving only the high spatial frequency components that contained anatomic information. For the multi-echo images, filtered phase maps for each TE were created separately and then combined using a weighted average with weights given by eq. 1.

The imaging plane used for MRTI was aligned with the MRI bore, not the ACPC plane (Fig. S1). The mean rotation about the lateromedial axis with respect to the ACPC plane was 4.8 ± 4.7°. The rotations about the dorsoventral and anteroposterior axes were 1.2 ± 2.1° and 0.6 ± 2.6°, respectfully.

### C. Segmentation

For each treatment, we segmented the RN in axial phase images and the STN and the RN in coronal phase images. For this segmentation, we selected the first sonications in each treatment in the different imaging orientations, and when available, for different frequency-encoding directions. The coordinates of the centroids of these segmentations were recorded. In addition, we segmented the internal capsule in both axial and coronal images. Finally, we segmented the central hyperintense region of the lesion in T2-weighted imaging obtained 24 h after treatment. When coronal T2-weighted images were not acquired at 24 h, we segmented the hypointense areas in T1-weighted images. These segmentations were performed on axial and coronal images interpolated to the planes used for MRTI after registration to the treatment planning images described below. We also recorded the locations of the sonication targets selected by the neurosurgeon and the centroid of the accumulated thermal dose. In the ACPC plane we recorded the left/right coordinate of the third ventricle.

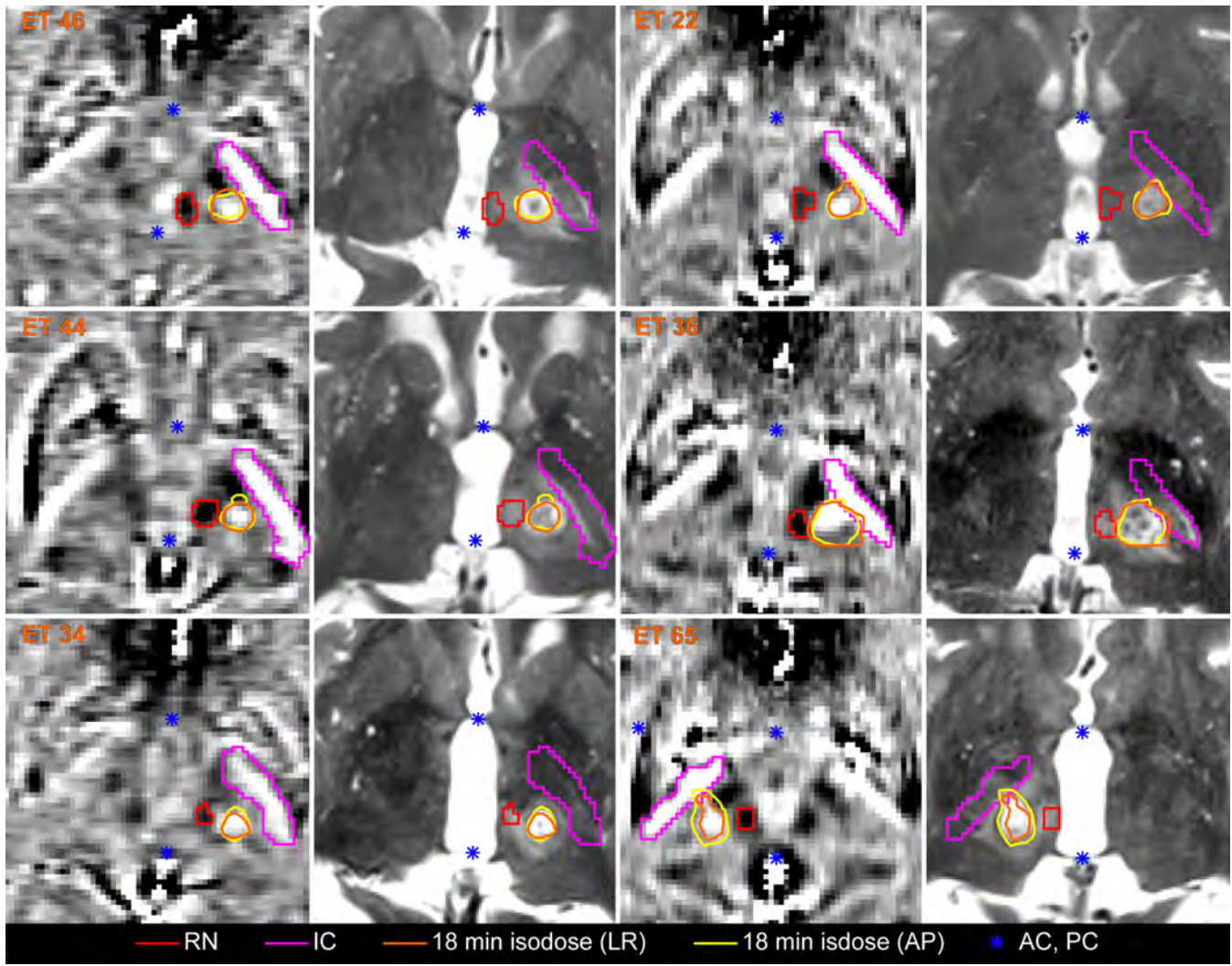

Fig. 3 Axial phase maps acquired during sonication (left) and corresponding T2-weighted images acquired 24 h later (right) for six treatments in ET patients. The T2-weighted images were registered to the treatment images. The segmentations and accumulated thermal dose contours superimposed on the T2-weighted images were all generated from the image data obtained during sonication. Dose contours were created separately for LR and AP frequency-encode directions in MRTI. The locations of the AC and PC are also shown.

### D. Image registration

Treatment planning MRI used a 3D FIESTA sequence for ET patients, and either the FIESTA sequence or T2-weighted imaging in PD patients. Axial and coronal T2-weighted MRI was obtained 24 hours after treatment with a fast spin echo sequence with a slice thickness of 2 mm in every patient. These images were manually registered to the treatment planning images using 3D Slicer (www.slicer.org). When coronal T2-weighted images were not available, we used 3D T1-weighted images. We interpolated the axial and coronal planning images and the imaging obtained 24 hours after treatment to the imaging planes used for MRTI in MATLAB. To correct for minor residual errors after the manual 3D registration, we applied additional 2D registration between the interpolated planning and post-treatment imaging using the "imregtform.m" function in MATLAB.

### E. Accumulated thermal dose

The total thermal dose at the sonication target accumulated over multiple sonications at different targets and exposure levels. The MRTI imaging plane varied between sonications, so we could not directly compare the total accumulated thermal dose to the post-treatment imaging. To estimate the accumulated dose, we created temperature map templates for the imaging orientations used during each treatment.

To create the templates, we first fit the heating and cooling trajectories of the hottest voxel to exponential functions:

$$\Delta T_{fit}(t) = (a_1 - e^{-a_2 t})e^{-a_3 t} \quad \text{(heating)}$$

$$\Delta T_{fit}(t) = b_1 e^{-b_2(t - t_d)} \quad \text{(cooling)}$$

The $a_n$ and $b_n$ coefficients were determined using the "nlinfit.m" function in MATLAB with robust options. If the time of the MRTI acquisition was within 1 s of the sonication duration ($t_d$), it was included in the heating curve fit even if it exceeded the sonication duration. The same rule was applied to the cooling. When extrapolated to the time of the heating duration, the fit heating and cooling curves often gave slightly different values. We scaled both curves by the average of the two curves at this time and re-normalized them. For each sonication, we fit ten time points to the heating and ten to the cooling.

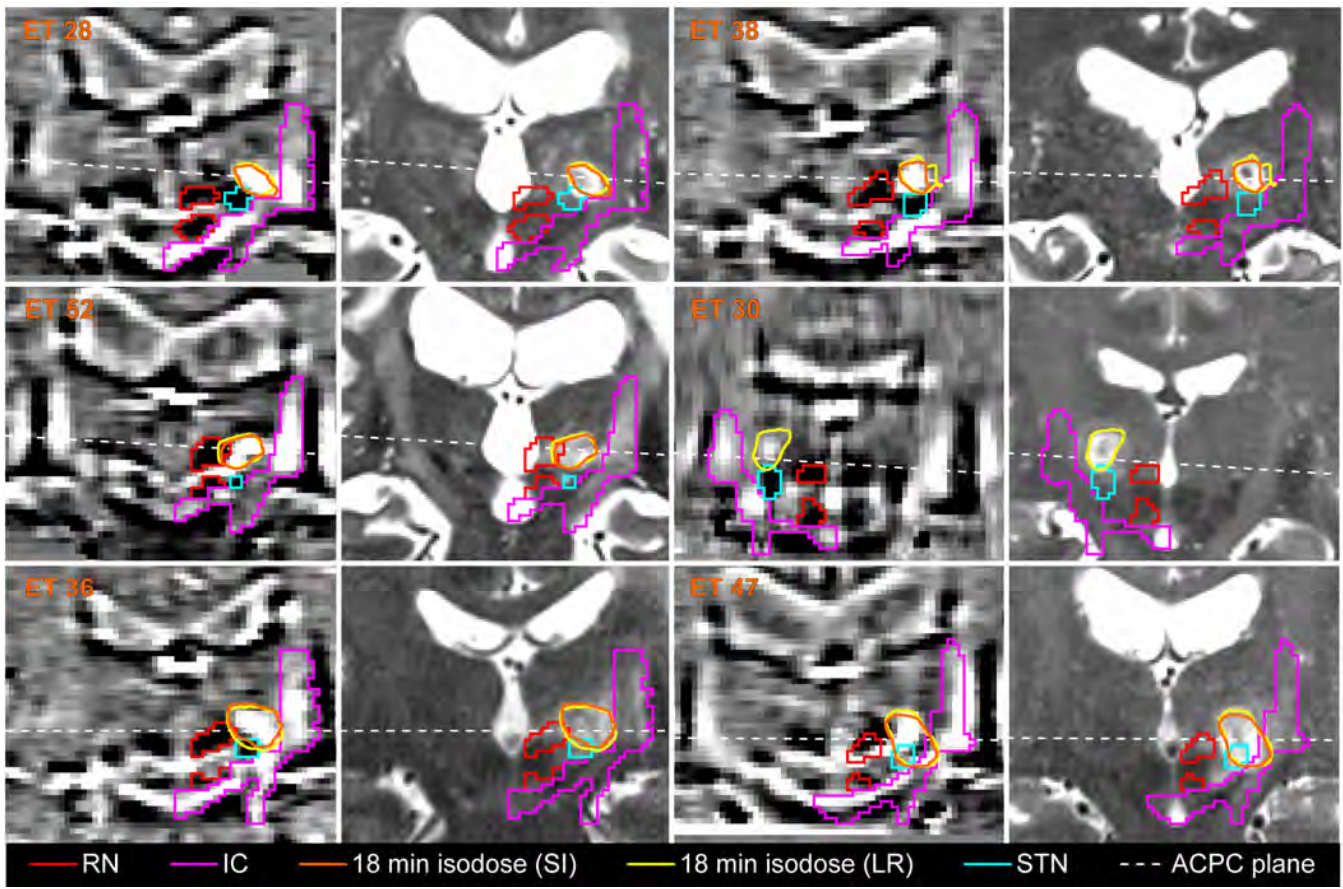

Fig. 4 Coronal phase maps acquired during sonication and corresponding T2-weighted images acquired 24 h later for six thalamotomies in ET patients. The T2-weighted images were registered to the treatment images. The segmentations and accumulated thermal dose contours superimposed on the T2-weighted images were all generated from the image data obtained during sonication. Dose contours were created separately for LR and SI frequency-encode directions in MRTI. The ACPC planes are also shown.

We next interpolated the MRTI spatially to a grid size of one fifth the native voxel dimensions and temporally to the same vector used in the exponential fits (ten time points during heating, ten during cooling). We then divided the MRTI by the maximum heating at the focus and multiplied them by $\Delta T_{fit}(t)/\max[\Delta T_{fit}(t)]$. For each imaging orientation, we performed a weighted average of these normalized data with weights determined by the peak temperature rise squared. We only included sonications where the peak temperature rise was greater than 7°C, and we created separate templates for different frequency-encoding directions.

We estimated the accumulated thermal dose in the axial and coronal planes used in MRTI in the first sonications of each treatment. To include sonications with different orientations, we fit the temperature rise as a function of time for the hottest voxel to exponential functions as described above and then multiplied it by the template to create simulated MRTI. If the focus was steered to a different coordinate within the plane of the template, we used interpolation to translate the template. When it was out of plane, we multiplied by a derating factor obtained from the spatial profile of the heating in the MRTI templates of that treatment.

The estimated accumulated thermal dose maps were used here to compare the location of the MRTI-estimated thermal dosimetry to anatomic structures in the filtered phase maps. Further details on the agreement of the accumulated thermal dose maps with the resulting lesions in MRI will be presented in a future report. Throughout this paper, thermal dose contours indicate 18 equivalent minutes at 43°C, a value previously found to be the threshold for thermal brain damage in rabbits [23].

### F. Data analysis

The main goal of this work was to determine the consistency of the detectability and the locations of identifiable anatomic structures in the phase mapping. This analysis was done primarily by manual inspection of the segmented structures and the registered post-treatment MRI. To further this goal, we combined the results of all treatments. The coordinates of the voxels within the segmentations of the RN, STN, and IC along with those within the accumulated thermal dose contours were rotated to a common reference frame so that the PC was located at the origin and the ACPC line was oriented along the anteroposterior direction. For visualization, this information was used to generate maps of the probability for each voxel to be within a segmentation of the RN, STN, or IC, or within the accumulated thermal dose contour. The coordinates of the targets, the lesions at 24 h, the centroids of the thermal dose

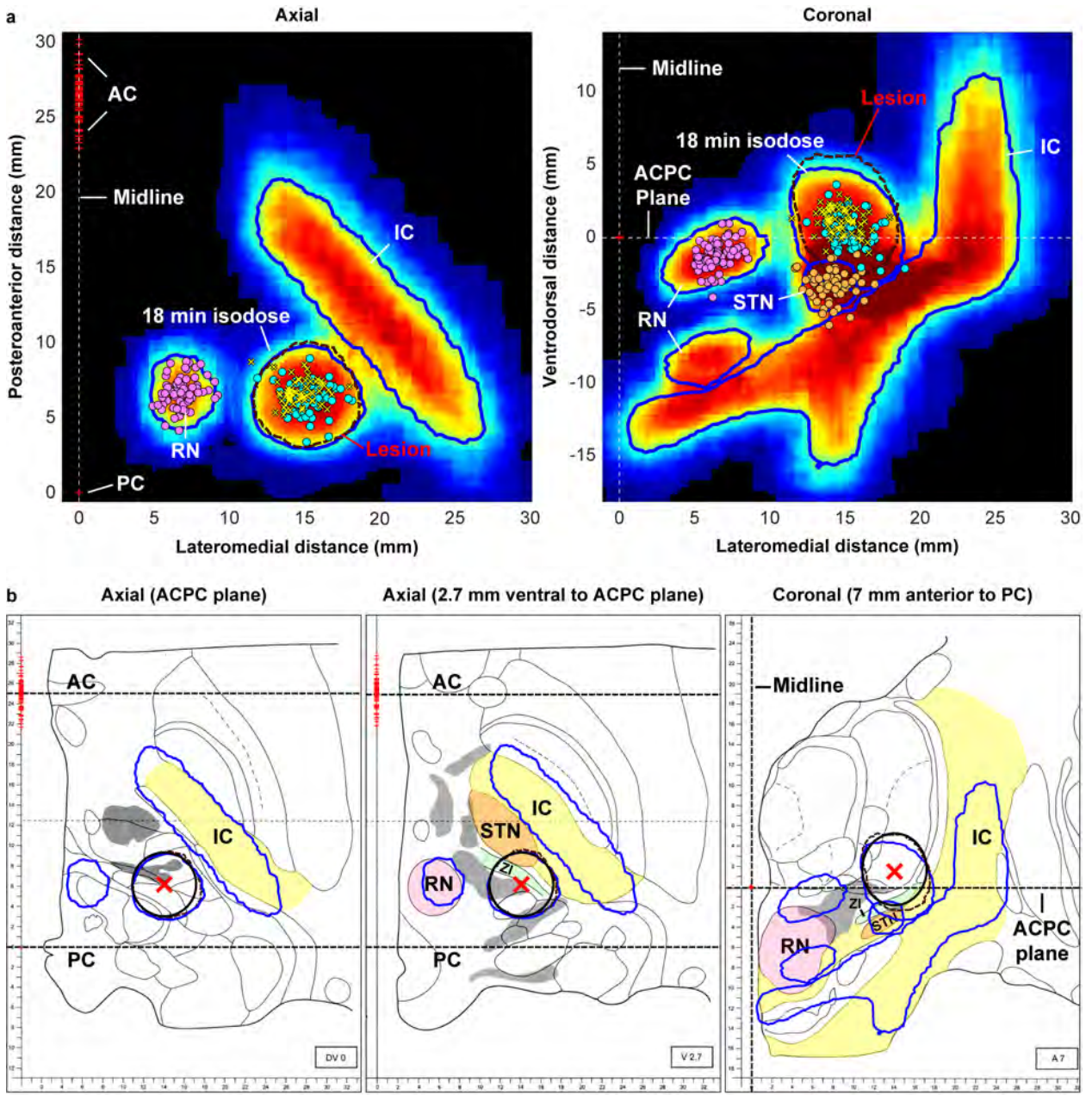

Fig. 5 (a) Target locations shown with respect to anatomic structures and thermal dose contours obtained with the phase maps. The images show the probability each point being within either a thermal dose contour or a segmentation of the RN, IC, or STN. The centroids of these segmentations (circles) along with the target coordinates (x) are also shown. The contours indicate regions where the probability for being within a segmentation or isodose line was 50%. The lesion contour (dashed line) indicates the 50% probability of being within the lesion segmentation in MRI obtained at 24 h. The maps were created after aligning the ACPC planes. (b) The contours from (a) superimposed on panels from the Morel atlas of the thalamus [1]. In the ACPC plane (DV0), the positions of the IC and the atlas-based target (red x) agreed on average. The RN is evident in an axial panel 2.7 mm ventral to the ACPC plane (V2.7) and aligned in the posteroanterior direction with the atlas target. In a coronal panel 7 mm anterior to the PC (A7), the locations of the STN in the atlas and in the phase maps overlapped, and the lateral positions of the RN were consistent. Overall, the targets were consistently positioned with respect to the RN in the AP direction, the STN in the LR direction, and ACPC plane and RN in the SI direction.

contours, and the centroids of the RN and STN segmentations were compared using paired t-tests. A value of P<0.05 was considered statistically significant.

## III. RESULTS

### G. ET treatments

Fig. 2 shows representative examples of axial and coronal phase maps from ET patients annotated with different anatomic structures. Segmentations of the IC and the RN are superimposed along with contours of the accumulated thermal dose delivered over the respective treatments. In coronal imaging, the RN was split into a dorsal and ventral component as described elsewhere [24]. In coronal imaging, the STN is also segmented.

Fig. 3 shows examples of axial phase maps and the corresponding T2-weighted imaging acquired 24 h after treatment after registration for six patients. The locations of the

| Lateromedial (coronal) | mean ± stdev. | $P^a$ | N |
|---|---|---|---|
| Midline to target (mm) | 14.5 ± 1.0 | - | 68 |
| Ventricle to target (mm) | 10.4 ± 0.9 | - | 68 |
| Ventricle to isodose (mm) | 10.7 ± 1.2 | 0.189 | 67 |
| Ventricle to lesion (mm) | 10.2 ± 1.5 | 0.188 | 68 |
| Ventricle to STN (mm) | 9.6 ± 1.1 | 0.500 | 65 |
| Posteroanterior (axial) | mean ± stdev. | $P^a$ | N |
| ACPC distance (mm) | 26.3 ± 1.6 | - | 68 |
| PC to target (%ACPC dist.) | 25.3 ± 2.5 | - | 68 |
| PC to target (mm) | 6.7 ± 0.8 | - | 68 |
| PC to isodose (mm) | 6.8 ± 1.0 | <0.001 | 67 |
| PC to lesion (mm) | 7.9 ± 0.7 | <0.001 | 68 |
| PC to RN (mm) | 7.2 ± 1.0 | 0.954 | 65 |
| Dorsoventral (coronal) | mean ± stdev. | $P^a$ | N |
| ACPC plane to target (mm) | 1.1 ± 0.8 | - | 68 |
| ACPC plane to isodose (mm) | 1.0 ± 1.2 | <0.001 | 68 |
| ACPC plane to lesion (mm) | 2.8 ± 1.1 | <0.001 | 68 |
| ACPC plane to RN$^b$ (mm) | -0.7 ± 0.9 | 0.051 | 68 |

[a]Compared to target distance with paired t-tests; lateromedial STN distance offset by 1 mm in the comparison; dorsoventral distance of RN offset by 2 mm; [b]dorsal component.

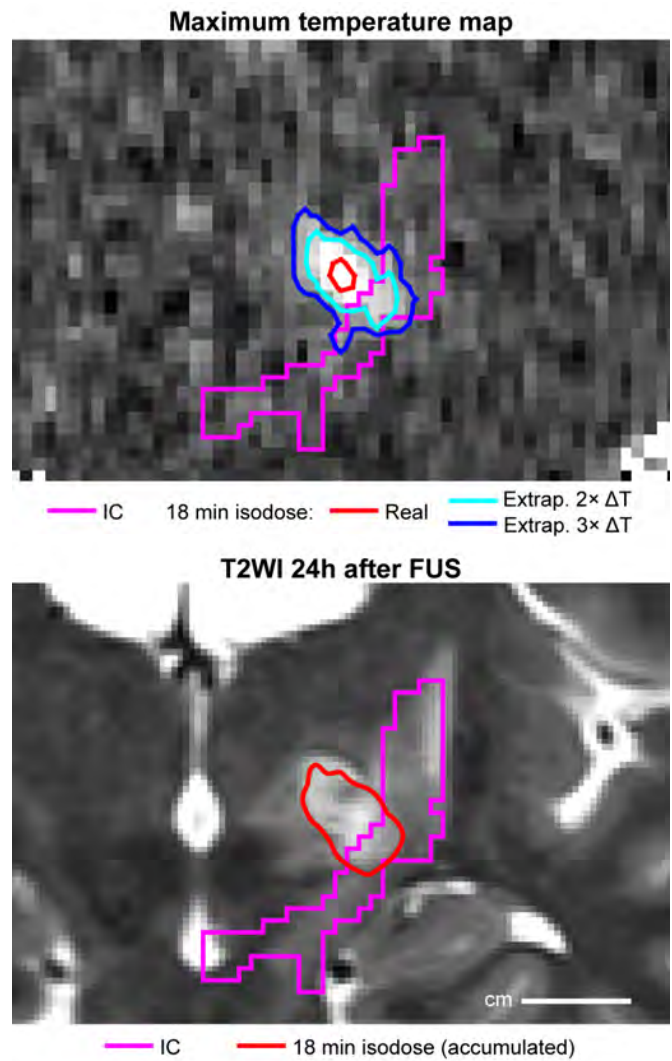

**Maximum temperature map**

IC — | 18 min isodose: | Real — | Extrap. 2× ΔT — | Extrap. 3× ΔT —

**T2WI 24h after FUS**

IC — | 18 min isodose (accumulated) —

Fig. 6 Potential use of this approach to avoid damage to the IC. The upper image is a map of the maximum temperature rise during a relatively low-energy sonication. The thermal dose estimated for this sonication (red contour) predicted only a small lesion that did not include the IC (magenta contour). Estimating what might occur if the temperature rise was increased by a factor of two or three (turquoise, blue contours) indicates that higher-energy sonication would damage the IC. Without this knowledge of the IC location, subsequent sonications were performed at higher energy that resulted in a lesion that included the IC (lower image). The accumulated thermal dose from all the sonications in the treatment superimposed on this image agrees reasonably with the extrapolation of the low-energy sonication on the left and the lesion evident in MRI at 24 h after MRgFUS.

RN, STN, and the IC and along with the thermal dose contours were all generated from the phase data used to create the MRTI. The IC was observed in all (68/68) treatments in axial imaging. The RN was detected in 96% (65/68) of the treatments in axial imaging. It was not seen in two treatments where the sonication target was localized superior to the RN. In another, artifacts caused by a previous DBS electrode obscured the RN.

Coronal examples for six patients are shown in Fig. 4. The IC was clearly observed in every patient. However, lateroventral portions of the IC (* in Fig. 2) were often not visible in coronal imaging, but they could be inferred by extrapolation. The STN was visualized in 96% (65/68) of the treatments; when it was not detected it was confounded by phase changes induced by heating. In some cases, heat-induced phase changes also overlapped with the IC, making discrimination between the two challenging.

Filtered phase images, and corresponding imaging acquired at 24 h after MRgFUS for all patients are included in the Supplemental Data. Inspection of these data demonstrate the consistent delineation of the IC, RN, and in coronal imaging, the STN. Sagittal imaging was less revealing, but often showed the anteroposterior trajectory of the STN (Fig. S2).

The IC, RN and STN were segmented in the coronal and axial phase imaging for 68 ET patients. The coordinates of the voxels contained within the segmentations along with those of the accumulated thermal dose isocontours were rotated to a common reference frame defined by the AC, the PC, and the midline. We used this information to visualize the average anatomic, dosimetry, and targeting information for the 68 patients. This analysis is summarized in Fig. 5, where the color map shows the relative probability of each point in the coronal and axial images of being within the segmentations or isodose contours. The locations of these regions and the sonication targets are also shown. For coronal imaging, we used the dorsal portion of the RN. Taken together, the targets and the thermal dose contour centroids were aligned with the RN anteroposteriorly. The targets were approximately 1 mm lateral to the STN lateromedially, 2 mm dorsal to the RN in coronal imaging. We also superimposed the 50% contour for the lesions segmented from imaging at 24 h after treatment.

Table 1 shows the distance between landmarks identified in the treatment planning imaging (AC, PC, midline, ventricle wall) and post-treatment imaging (lesion) to those obtained from phase mapping obtained during sonication (the centroids of the RN/STN, accumulated thermal dose contours). As expected, the targets and isodose centroids were, on average, close to the nominal atlas-based coordinates of 11 mm lateral to the ventricle wall, anterior to the PC at 25% of the ACPC distance, and 1 mm dorsal to the ACPC plane. The variability in these coordinates – a standard deviation of approximately ±1

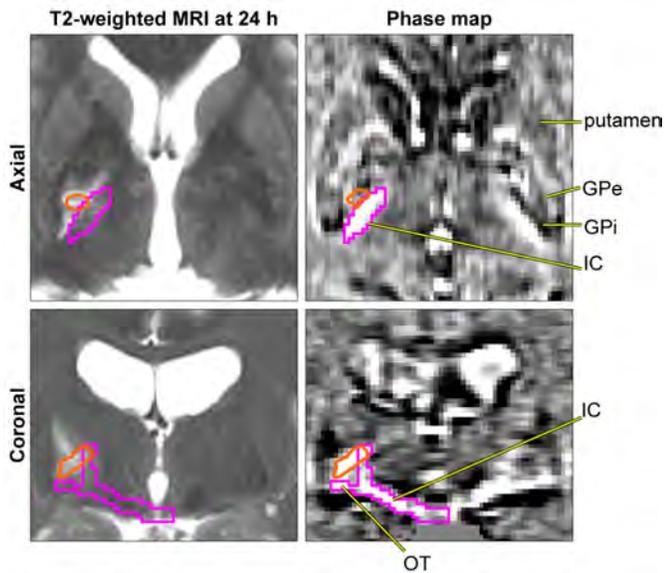

**T2-weighted MRI at 24 h**     **Phase map**

Axial

Coronal

putamen

GPe

GPi

IC

IC

OT

Fig. 7 Phase maps and corresponding T2-weighted images from a pallidotomy in a patient with Parkinson's disease. The IC was visible in the phase maps in both orientations. The GPi appeared hypointense in axial phase images, but was not as readily discerned in the coronal images. The optic tract was observed ventral to the GPi target in coronal imaging.

mm – was similar to that found for the centroids of the RN and STN. The differences between the centroids of the RN anteroposteriorly, the STN offset medially by 1 mm medially, and the RN offset 2 mm dorsally and the target coordinates were not statistically significant (P>0.05).

Using both the anatomic and thermal information together could provide a way to extrapolate low energy sonications and assess the risk to the IC. An example of how this might be used is shown in Fig. 6, where a sonication obtained at a relatively low energy is scaled to predict the thermal dose at a higher energy [23]. In the patient shown in Fig. 6, without knowledge of the location of the IC, subsequent sonications at higher energy were delivered that resulted in IC damage.

### A. PD treatments

Fig. 7 shows example images from a pallidotomy in a PD patient. As was the case in ET patients, the IC was visible in axial imaging. Axial imaging also revealed the GPi as a hypointense area next to the IC. Coronal phase imaging allowed for visualization of the IC medial to the sonication target. A hyperintense area was also observed ventral to the sonicated target with a location consistent with the optic tract. This structure was also evident in sagittal imaging (Fig. S2). Imaging from the three additional pallidotomies are included in the Supplemental Data.

### IV. DISCUSSION

Anatomic structures in the filtered phase maps were consistently visualized with high image contrast. These maps were created using the same imaging data used to generate the MRTI. While they had a low spatial resolution, they were automatically registered to the MRTI and included any image distortion that may have occurred when changing the imaging orientation and frequency-encoding direction (supplemental video 1). This distortion can be a millimeter or more with the low-bandwidth, single-echo MRTI sequence. Furthermore, the phase maps require no extra time to acquire. Such maps could also be acquired separately from the sonication at higher resolution if necessary.

The IC was clearly visualized in every patient in the filtered phase maps. Protecting the IC is critical during MRgFUS thalamotomy and pallidotomy to avoid hemiparesis or other side effects [2]. Risk to the IC is exacerbated by the obliquity of the heating that is often observed [8], the proximity of the VIM and GPi to this tract, and the lower blood flow in dense white matter fiber tracts which may retard heat dissipation.

In an earlier study in rabbits [23], we showed that one can use MRTI acquired during low-energy sonication to accurately predict the thermal dose at subsequent high-energy exposures. Using this approach along with visualization of the IC could prove useful to avoid damage (Fig. 6). Cognizant of the proximity to the IC, one could logically reposition the target more medially and/or dorsally in such cases. It might also be possible to use the accumulated thermal dose of multiple low-energy sonications [25] at different closely-spaced locations to avoid damage while still achieving a thermally-ablated lesion with a sufficient volume.

Along with safety concerns, the anatomic information in the phase maps might be useful to refine atlas-based targeting. The MRTI was centered on the target selected by the neurosurgeon. This target was determined based on anatomic measurements and adjusted based on symptom improvement, which was achieved in every patient. The agreement in the alignment of the target anteroposteriorly with the RN and the proximity of the lesion boundary to the IC laterally and the STN ventrally in both stereotactic atlases [1] and in the phase mapping suggest that these structures could also be used as landmarks to refine the standard targeting used by most neurosurgeons. Future investigation with imaging that can directly visualize the VIM [26, 27] or that visualize relevant white matter tracts with diffusion tensor imaging [28, 29] might be illuminating in understanding how direct targeting approaches compares to atlas-based targeting refined by anatomic landmarks visualized in the phase maps. While such direct targeting approaches are likely the gold standard in targeting for thalamotomy, they require non-standard imaging approaches, long scan times, and advanced image processing methods to visualize the VIM that may not be available to all sites. Furthermore, the need to acquire a pre-treatment MRI is an added expense.

In PD patients, the ability to visualize the optic tract ventral to the sonication target in addition to the IC should also be useful for avoiding side effects. It also appears that the phase imaging can delineate the GPi, presumably due it its high iron content [30]. However, only four patients were examined, and additional work is needed to verify the consistency of the phase imaging to guide pallidotomy.

It is unfortunate that a gap was consistently apparent in the coronal phase maps in lateroventral regions of the IC in many ET patients, as this area is at risk for off-target damage when the focal area is oblique. The reason for this gap is unclear, but it could be related to the angle of the white matter tracts fibers with respect to the magnetic field in those regions. The MRI magnitude and phase in organized tissue structures such as white matter tracts are orientation-dependent due to dipolar interactions between bound water protons. This effect has a null value when the structure is at the "magic angle" of 54.74° with respect to the magnetic field [4], which is consistent with our observations. The location of the IC in this gap could be inferred by extrapolation, but its precise boundary was less clear than in other areas.

Another limitation of this study is that the MRTI was not aligned with the ACPC plane, but rather varied slightly between subjects. If this variability makes image interpretation challenging, one could separately acquire images aligned with the patient instead of the scanner. Also, the phase changes induced by heating often overlapped with the anatomic information. In some patients, this phase was sufficient to obscure the STN and IC in coronal imaging. This issue could be avoided by acquiring the images separately without sonication. It may be possible to use the temperature estimates to remove the heating-induced phase changes from the anatomic visualization.

### A. Conclusions

The image data used to create MRTI during MRgFUS thermal ablation contains anatomic information that is relevant for refining lesion placement in atlas-based targeting of the VIM and preventing unwanted damage to the IC. This anatomic information is automatically registered to the thermal mapping, does not suffer from uncertainty due image distortion, and requires no extra time to acquire. This approach could be useful in improving efficacy, safety and treatment time during this noninvasive neurosurgical procedure.

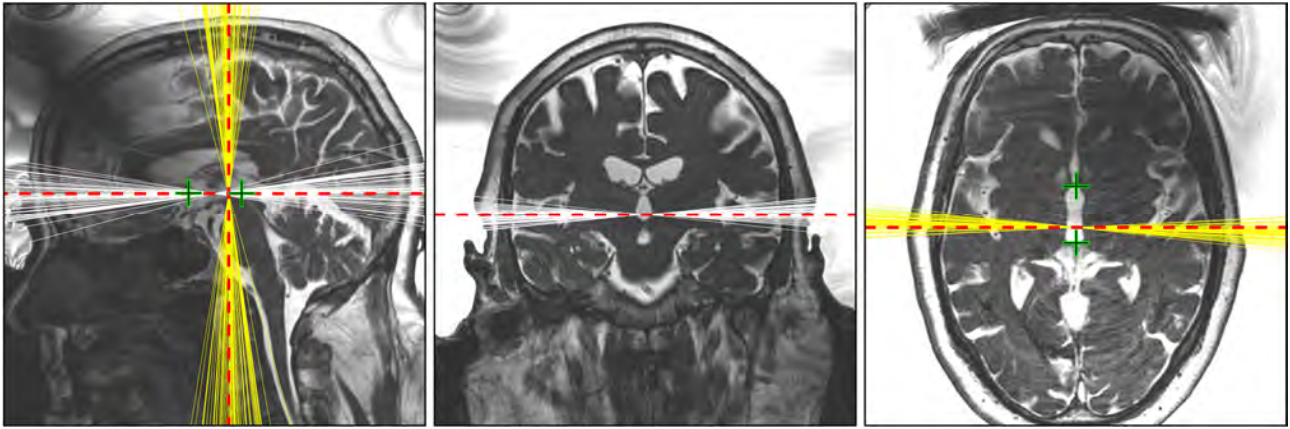

Fig. S1 Angulation of the planes used for MRTI and phase mapping with respect to the ACPC plane. The MRTI was aligned with the MRI scanner, so its orientation varied between patients depending on the placement of the stereotactic frame. To visualize the tilt angle, the angulation of the planes used for coronal and axial MRTI in the 68 ET treatments are superimposed on the FIESTA treatment planning images from one patient. The AC and PC, the ACPC plane, and the plane perpendicular to the ACPC plane at 25% of the ACPC distance are also shown.

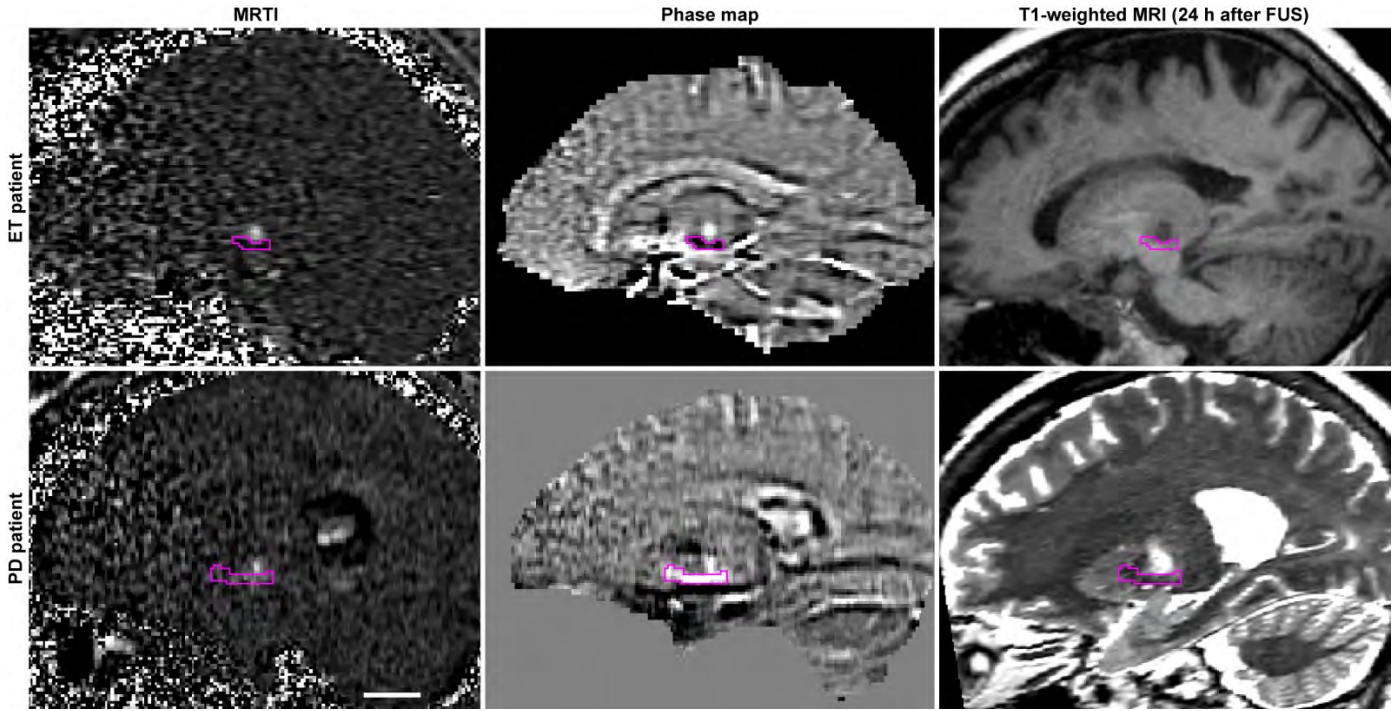

Fig. S2 Sagittal examples for a thalamotomy in an ET patient (top) and pallidotomy in a patient with PD (bottom). In ET patients, the STN was hypointense and appeared directly below the heated region in the phase mapping. In PD patients, the optic tract was directly below the heated region and was hyperintense. T1-weighted MRI acquired at 24 h after FUS are shown. In the PD patient, the FGATIR sequence was used (Sudhyadhom et al., Neuroimage, 7; T44-T52, 2009).

| Patient | | Age | # soni-cations | Multi-echo MRTI | ACPC dis-tance (mm) | Midline to target (mm) | Ventricle to target (mm) | PC to target (%ACPC dist.) | PC to target (mm) | ACPC plane to tar-get (mm) |
|---|---|---|---|---|---|---|---|---|---|---|
| ET | 1 | 58M | 15 | N | 24.7 | 11.4 | 14.0 | 24.2 | 6.0 | -1.0 |
| ET | 2 | 65F | 20 | N | 24.8 | 11.0 | 14.0 | 22.1 | 5.5 | 0.3 |
| ET | 3 | 78M | 17 | N | 27.5 | 11.7 | 15.3 | 27.0 | 7.4 | -0.2 |
| ET | 4 | 72F | 15 | N | 25.0 | 10.8 | 13.9 | 24.6 | 6.2 | 0.1 |
| ET | 5 | 65M | 13 | N | 26.4 | 11.4 | 14.9 | 25.5 | 6.7 | 0.4 |
| ET | 6 | 69M | 28 | N | 27.4 | 11.9 | 15.8 | 20.4 | 5.6 | -0.3 |
| ET | 7 | 68M | 15 | N | 28.2 | 10.8 | 14.8 | 19.6 | 5.5 | 0.7 |
| ET | 8 | 77F | 17 | N | 24.9 | 10.8 | 13.8 | 22.8 | 5.7 | 0.1 |
| ET | 9 | 83F | 12 | N | 24.1 | 10.6 | 14.3 | 24.5 | 5.9 | 0.3 |
| ET | 10 | 80M | 12 | N | 29.2 | 10.1 | 16.5 | 23.3 | 6.8 | -1.1 |
| ET | 11 | 70M | 9 | N | 28.2 | 9.5 | 14.8 | 22.7 | 6.4 | -0.1 |
| ET | 12 | 80F | 12 | N | 23.5 | 10.9 | 15.1 | 25.1 | 5.9 | 0.3 |
| ET | 13 | 91M | 10 | N | 26.8 | 9.9 | 16.0 | 21.5 | 5.8 | 0.3 |
| ET | 14 | 71M | 13 | Y | 27.7 | 9.9 | 13.9 | 22.9 | 6.3 | 1.0 |
| ET | 15 | 75M | 12 | Y | 26.4 | 10.7 | 14.0 | 23.8 | 6.3 | 1.6 |
| ET | 16 | 78M | 20 | Y | 27.6 | 9.6 | 14.3 | 23.9 | 6.6 | 1.2 |
| ET | 17 | 79F | 12 | Y | 25.4 | 10.4 | 13.8 | 26.2 | 6.6 | 1.1 |
| ET | 18 | 71M | 16 | Y | 26.3 | 12.4 | 15.2 | 27.3 | 7.2 | 1.0 |
| ET | 19 | 69F | 13 | Y | 23.3 | 11.2 | 13.8 | 23.2 | 5.4 | 1.3 |
| ET | 20 | 84M | 10 | Y | 26.3 | 11.7 | 17.0 | 26.5 | 7.0 | 1.3 |
| ET | 21 | 77M | 10 | Y | 25.8 | 11.4 | 13.5 | 23.6 | 6.1 | 1.7 |
| ET | 22 | 86F | 10 | Y | 25.4 | 11.0 | 14.2 | 25.1 | 6.4 | 1.8 |
| ET | 23 | 72M | 12 | Y | 23.7 | 10.5 | 13.7 | 23.1 | 5.5 | 1.6 |
| ET | 24 | 84M | 13 | Y | 27.3 | 10.6 | 15.2 | 25.4 | 6.9 | 1.0 |
| ET | 25 | 72F | 11 | Y | 24.9 | 10.2 | 14.5 | 24.4 | 6.1 | 1.0 |
| ET | 26 | 81M | 13 | Y | 26.5 | 10.6 | 15.5 | 25.1 | 6.7 | 0.2 |
| ET | 27 | 79M | 12 | Y | 28.4 | 11.9 | 15.4 | 26.2 | 7.5 | 1.9 |
| ET | 28 | 83M | 12 | Y | 30.1 | 10.2 | 16.1 | 23.4 | 7.1 | 1.0 |
| ET | 29 | 73M | 11 | Y | 27.0 | 10.8 | 14.2 | 25.4 | 6.9 | 0.4 |
| ET | 30 | 67F | 9 | Y | 24.0 | 11.6 | 13.5 | 25.6 | 6.2 | 2.2 |
| ET | 31 | 85F | 16 | Y | 26.1 | 10.6 | 13.6 | 20.2 | 5.3 | 1.7 |
| ET | 32 | 72M | 14 | Y | 26.8 | 11.3 | 14.8 | 28.4 | 7.6 | -0.1 |
| ET | 33 | 68M | 10 | Y | 25.7 | 10.7 | 14.6 | 26.0 | 6.7 | 0.6 |
| ET | 34 | 81M | 8 | Y | 28.7 | 8.9 | 13.9 | 24.0 | 6.9 | 1.9 |
| ET | 35 | 72F | 10 | Y | 27.0 | 10.5 | 16.6 | 25.3 | 6.8 | 1.5 |
| ET | 36 | 67M | 15 | Y | 26.9 | 10.4 | 14.6 | 25.5 | 6.9 | 1.0 |
| ET | 37 | 74M | 11 | Y | 26.6 | 9.7 | 14.7 | 23.4 | 6.2 | 0.8 |
| ET | 38 | 76F | 5 | Y | 22.9 | 9.4 | 13.1 | 23.3 | 5.3 | 1.6 |
| ET | 39 | 66M | 12 | Y | 25.0 | 11.8 | 14.2 | 30.4 | 7.6 | 1.4 |
| ET | 40 | 75M | 8 | Y | 27.7 | 10.2 | 15.1 | 25.4 | 7.0 | 1.6 |

| Patient | | Age | # soni-cations | Multi-echo MRTI | ACPC distance (mm) | Midline to target (mm) | Ventricle to target (mm) | PC to target (%ACPC dist.) | PC to target (mm) | ACPC plane to target (mm) |
|---|---|---|---|---|---|---|---|---|---|---|
| ET | 41 | 60F | 10 | Y | 26.1 | 10.9 | 14.0 | 23.4 | 6.1 | 1.8 |
| ET | 42 | 78M | 17 | Y | 25.6 | 9.3 | 14.0 | 23.9 | 6.1 | 1.7 |
| ET | 43 | 72M | 12 | Y | 24.0 | 11.1 | 15.7 | 27.2 | 6.5 | 0.4 |
| ET | 44 | 68F | 7 | Y | 26.1 | 10.0 | 13.6 | 25.6 | 6.7 | 0.7 |
| ET | 45 | 73F | 8 | Y | 23.7 | 11.6 | 14.1 | 26.2 | 6.2 | -0.4 |
| ET | 46 | 89F | 10 | Y | 24.1 | 8.8 | 14.4 | 25.5 | 6.2 | 0.6 |
| ET | 47 | 70F | 12 | Y | 25.8 | 10.6 | 13.3 | 25.5 | 6.6 | 1.7 |
| ET | 48 | 68F | 10 | Y | 25.9 | 10.7 | 14.4 | 28.0 | 7.2 | 2.1 |
| ET | 49 | 75F | 10 | Y | 27.8 | 9.5 | 14.7 | 30.6 | 8.5 | 1.8 |
| ET | 50 | 84M | 12 | Y | 27.1 | 8.9 | 13.5 | 24.2 | 6.6 | 2.2 |
| ET | 51 | 73F | 12 | Y | 26.1 | 11.1 | 14.1 | 31.5 | 8.2 | 2.5 |
| ET | 52 | 73F | 12 | Y | 25.9 | 6.7 | 12.5 | 32.3 | 8.4 | 0.0 |
| ET | 53 | 76M | 12 | Y | 26.7 | 8.2 | 11.5 | 32.2 | 8.6 | 1.1 |
| ET | 54 | 68M | 11 | N | 26.6 | 10.3 | 16.2 | 27.2 | 7.2 | 1.4 |
| ET | 55 | 79M | 12 | Y | 28.2 | 9.7 | 14.7 | 26.1 | 7.4 | 2.8 |
| ET | 56 | 79F | 11 | Y | 27.1 | 10.0 | 15.2 | 23.7 | 6.4 | 1.2 |
| ET | 57 | 66F | 7 | Y | 24.6 | 10.7 | 14.0 | 27.4 | 6.7 | 1.0 |
| ET | 58 | 65M | 10 | Y | 29.8 | 10.9 | 14.7 | 25.4 | 7.6 | 1.7 |
| ET | 59 | 87M | 8 | Y | 25.0 | 9.2 | 14.0 | 25.1 | 6.3 | 2.2 |
| ET | 60 | 83M | 6 | Y | 29.1 | 9.9 | 14.1 | 27.5 | 8.0 | 2.0 |
| ET | 61 | 92F | 15 | Y | 27.7 | 9.9 | 15.0 | 26.4 | 7.3 | 1.9 |
| ET | 62 | 87M | 8 | Y | 26.6 | 9.9 | 15.8 | 27.7 | 7.4 | 1.9 |
| ET | 63 | 74M | 12 | Y | 26.2 | 10.5 | 14.9 | 23.9 | 6.3 | 1.7 |
| ET | 64 | 73M | 8 | Y | 26.8 | 11.0 | 13.2 | 27.0 | 7.2 | 0.9 |
| ET | 65 | 79F | 10 | Y | 26.5 | 10.4 | 14.4 | 25.5 | 6.8 | 1.1 |
| ET | 66 | 89F | 8 | Y | 26.8 | 9.9 | 15.4 | 23.1 | 6.2 | 0.9 |
| ET | 67 | 89M | 9 | Y | 27.3 | 9.8 | 14.2 | 24.3 | 6.6 | 1.4 |
| ET | 68 | 69M | 9 | Y | 27.2 | 10.2 | 13.9 | 25.3 | 6.9 | 1.9 |
| PD | 1 | 73M | 24 | N | 26.1 | | | | | |
| PD | 2 | 48M | 13 | N | 26.2 | | | | | |
| PD | 3 | 41F | 10 | N | 25.2 | | | | | |
| PD | 4 | 61F | 7 | Y | 25.4 | | | | | |

**ET Patients 1-36**
**Axial T2WI; 24h after MRgFUS**

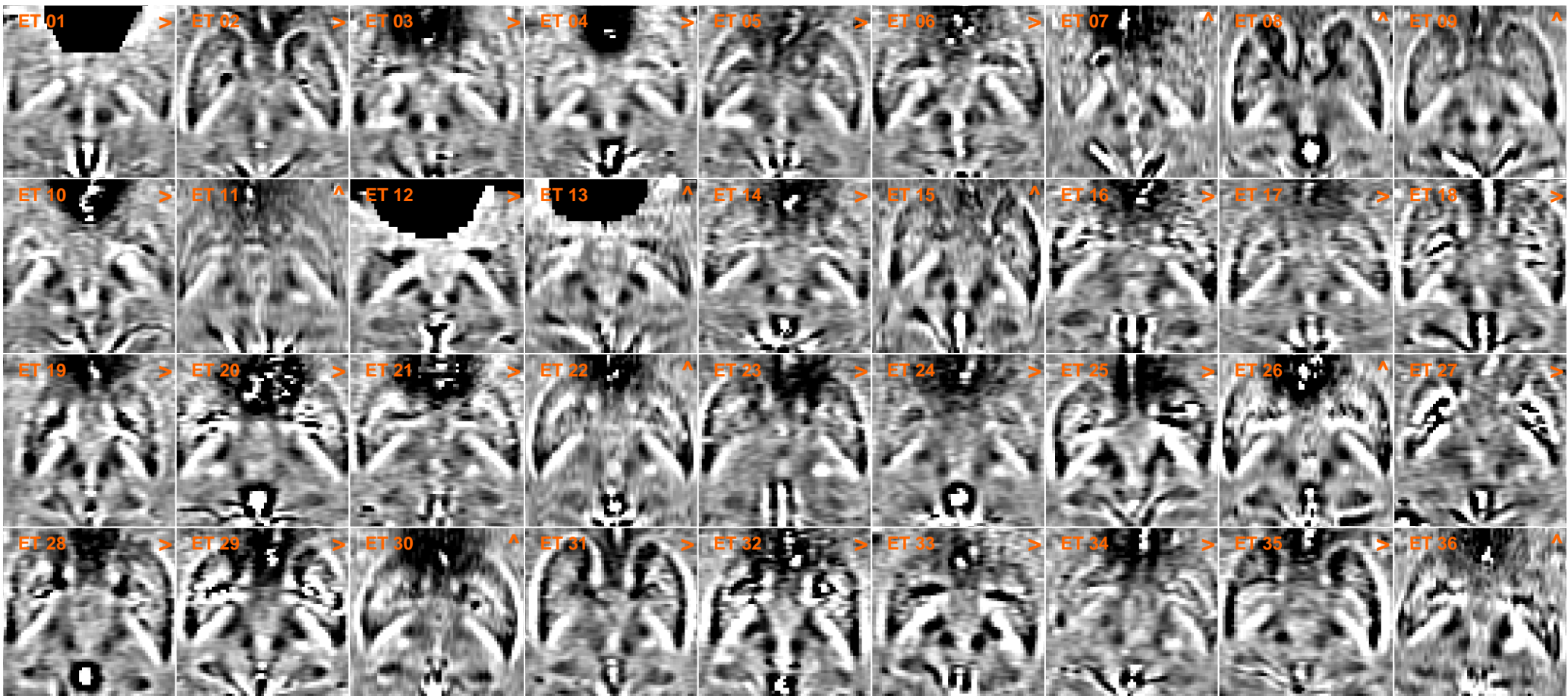

**ET Patients 1-36**
**Axial T2WI; 24h after MRgFUS**

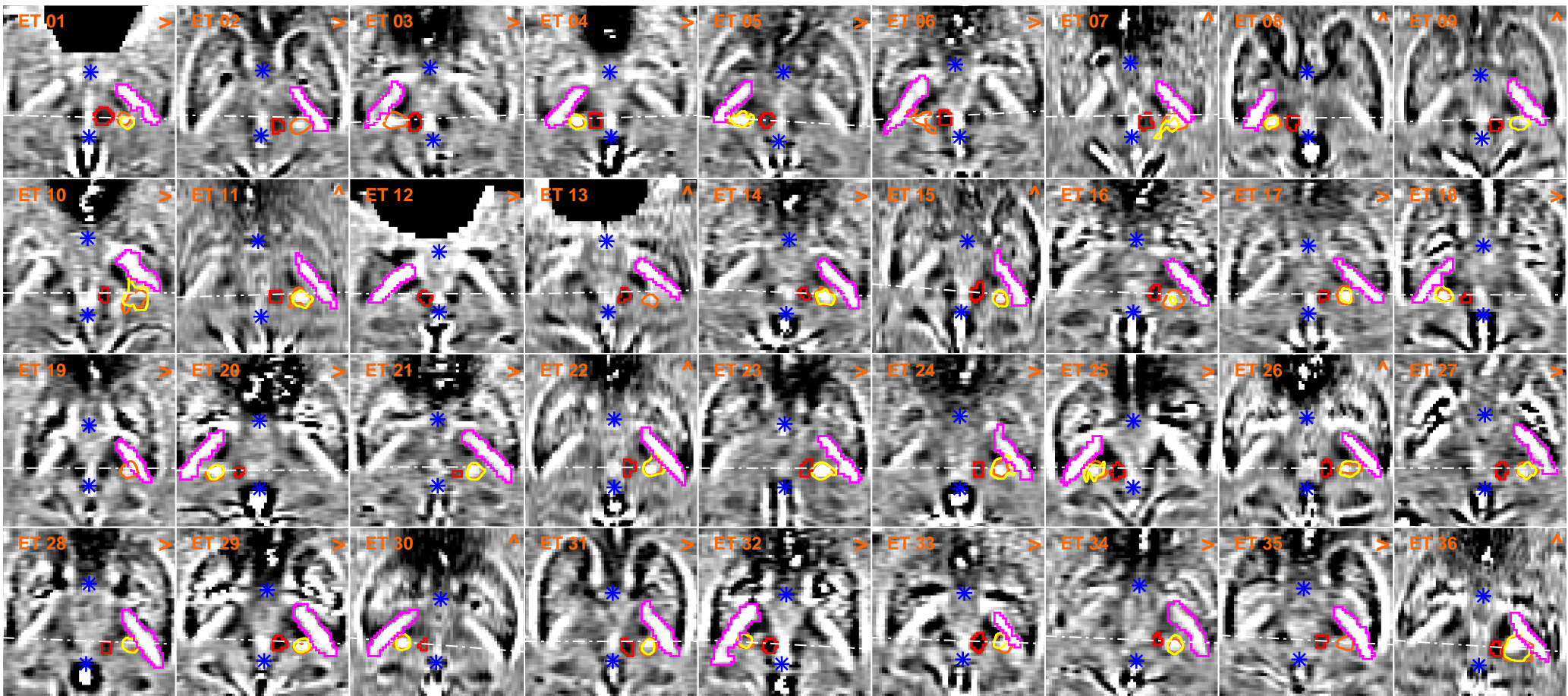

RN —  IC —  dose (AP) —  dose (LR) —  25% ACPC dist. - - -  AC, PC ✱

**ET Patients 1-36**
**Axial T2WI; 24h after MRgFUS**

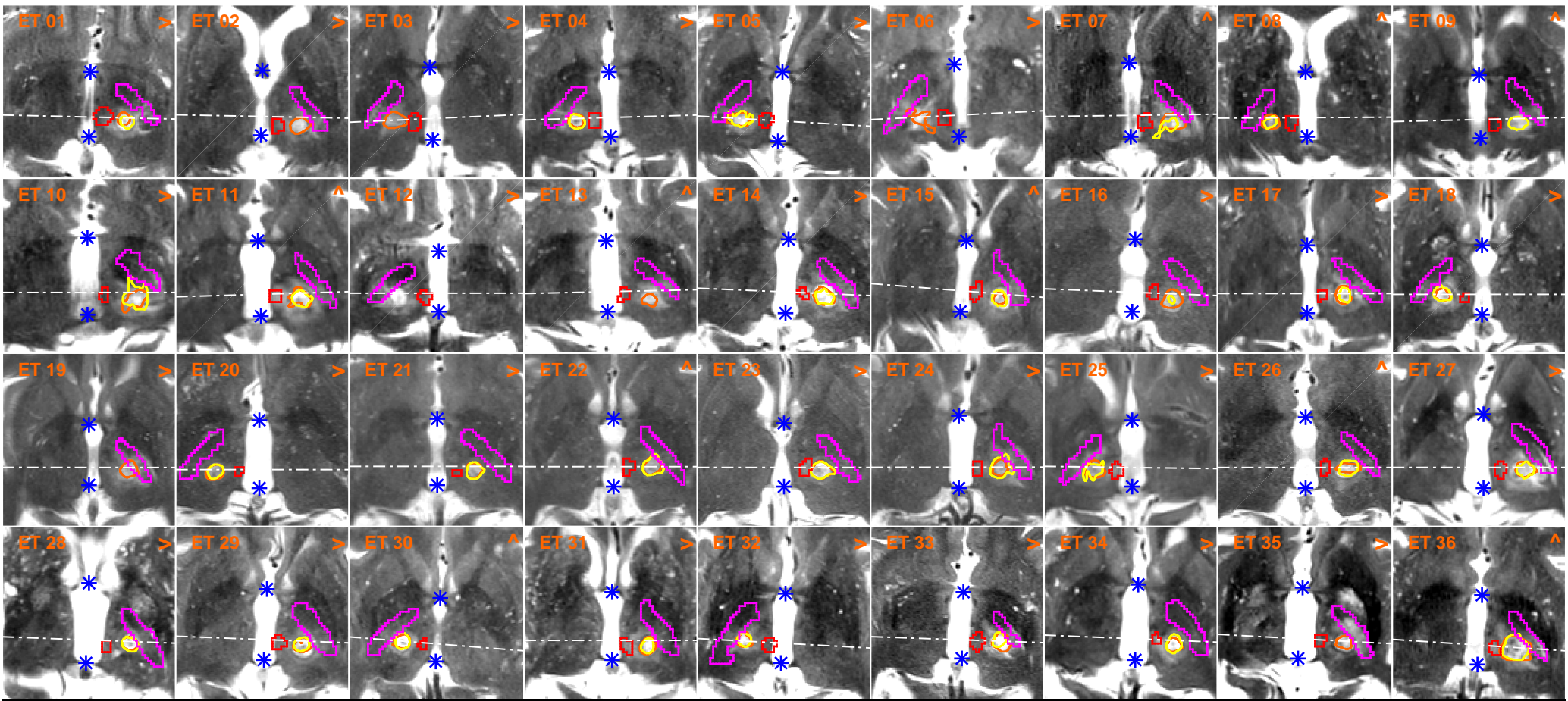

RN ———    IC ———    dose (AP) ———    dose (LR) ———    25% ACPC dist. –·–·–    ✳ AC, PC

**ET Patients 1-36**
**Axial T2WI; 24h after MRgFUS**

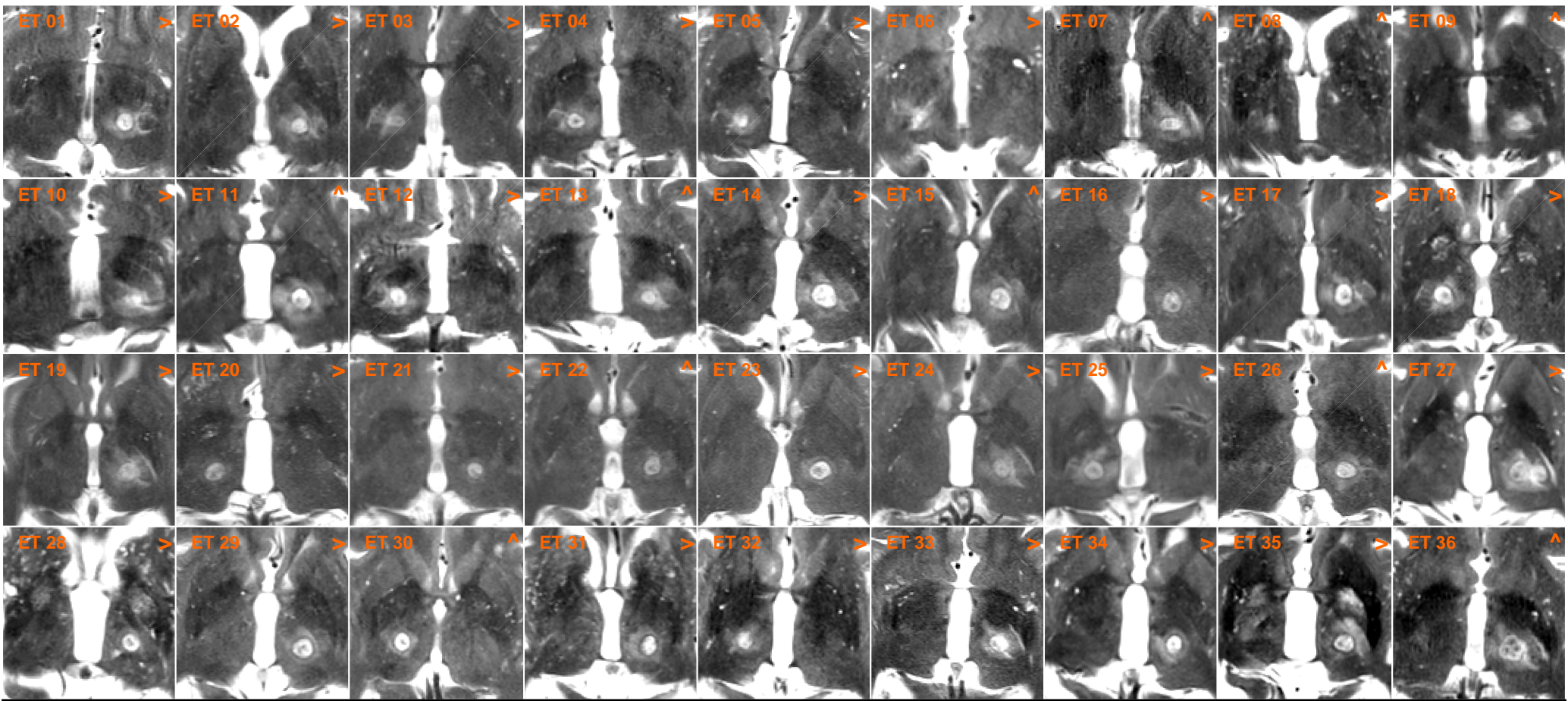

**ET Patients 37-68**
**Axial T2WI; 24h after MRgFUS**

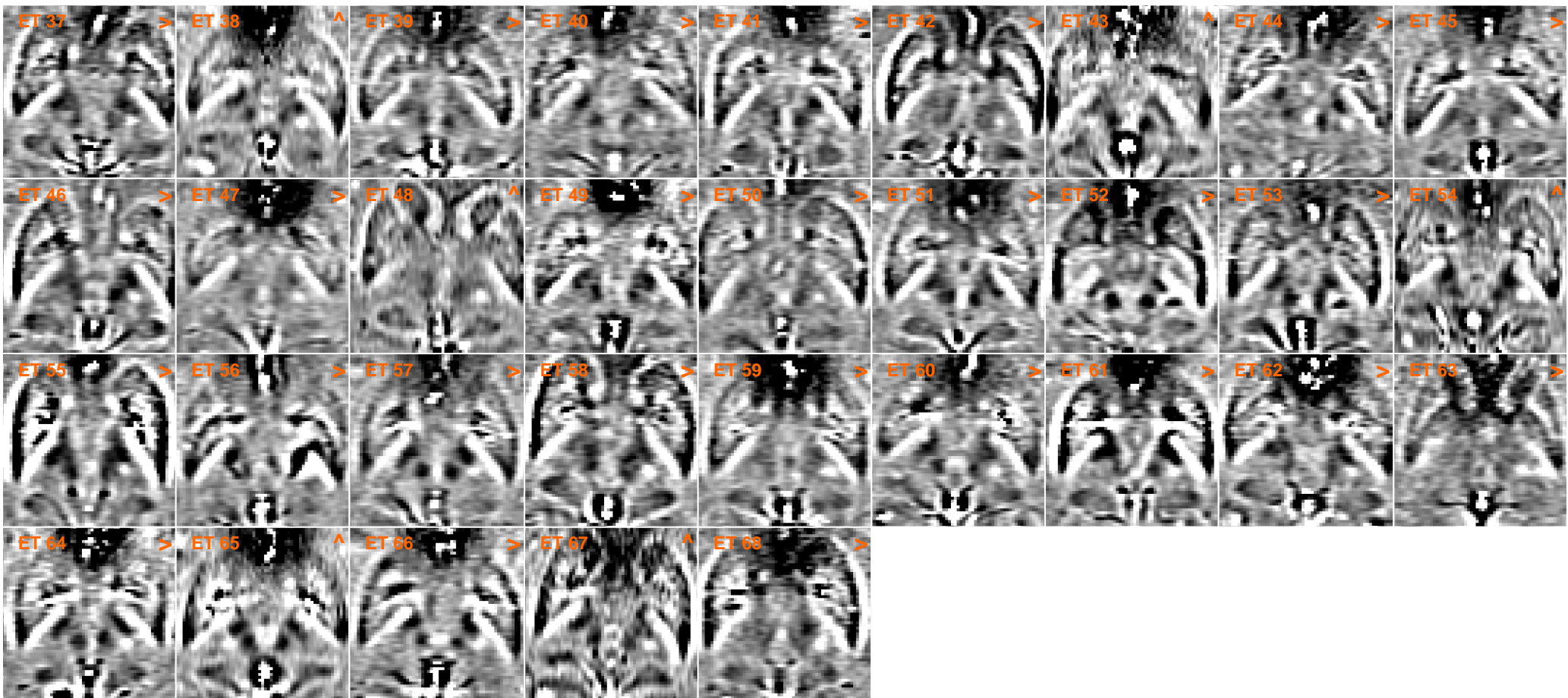

**ET Patients 37-68**
**Axial T2WI; 24h after MRgFUS**

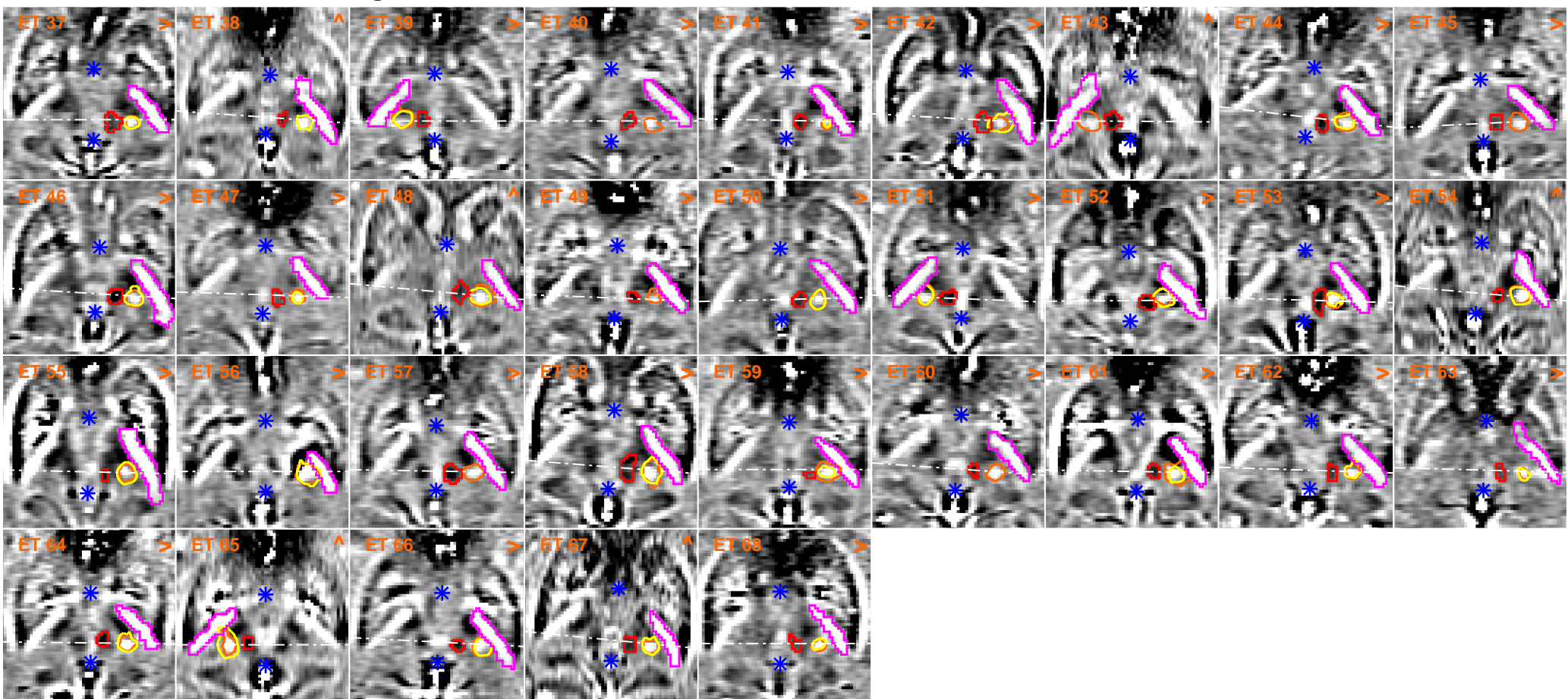

RN ━━━  IC ━━━  dose (AP) ━━━  dose (LR) ━━━  ┄┄┄ 25% ACPC dist.  ✻ AC, PC

**ET Patients 37-68**
**Axial T2WI; 24h after MRgFUS**

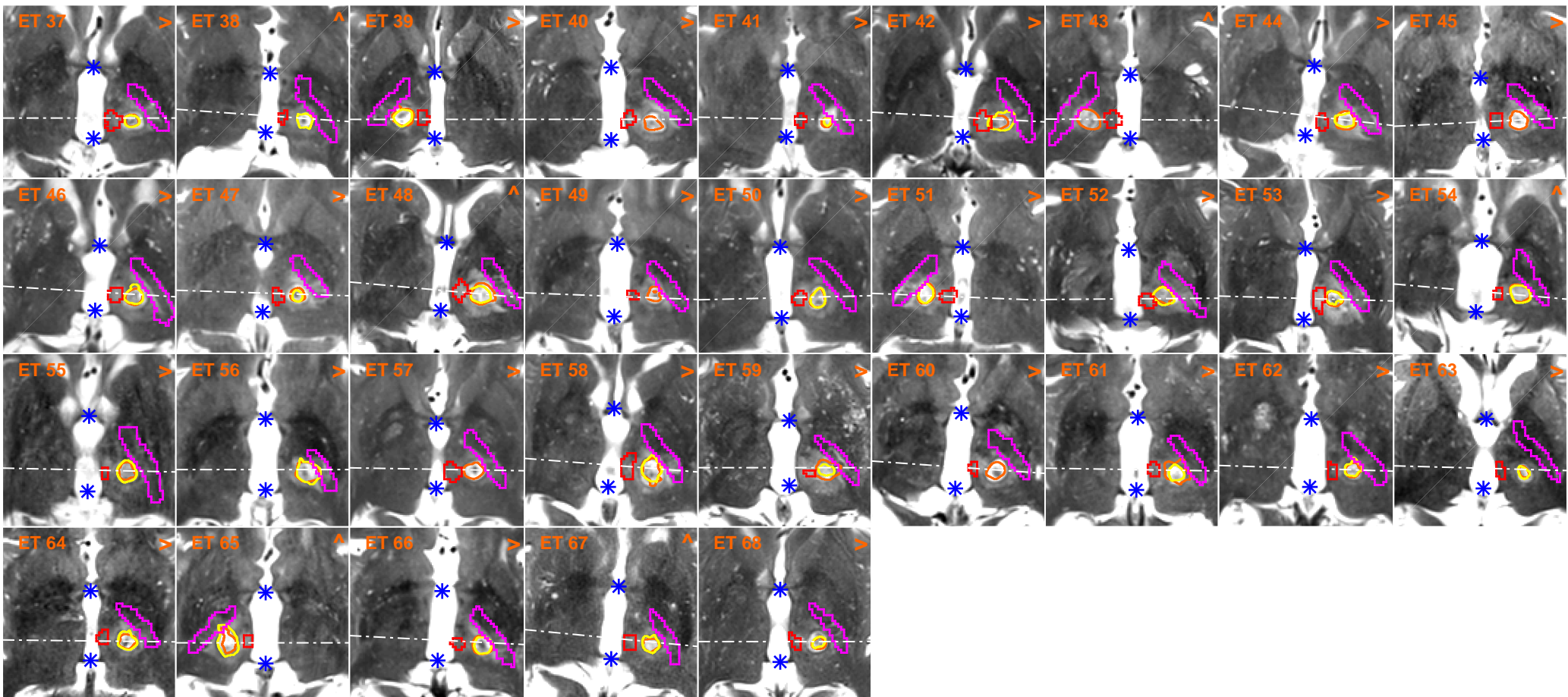

RN — IC — dose (AP) — dose (LR) — 25% ACPC dist. — * AC, PC

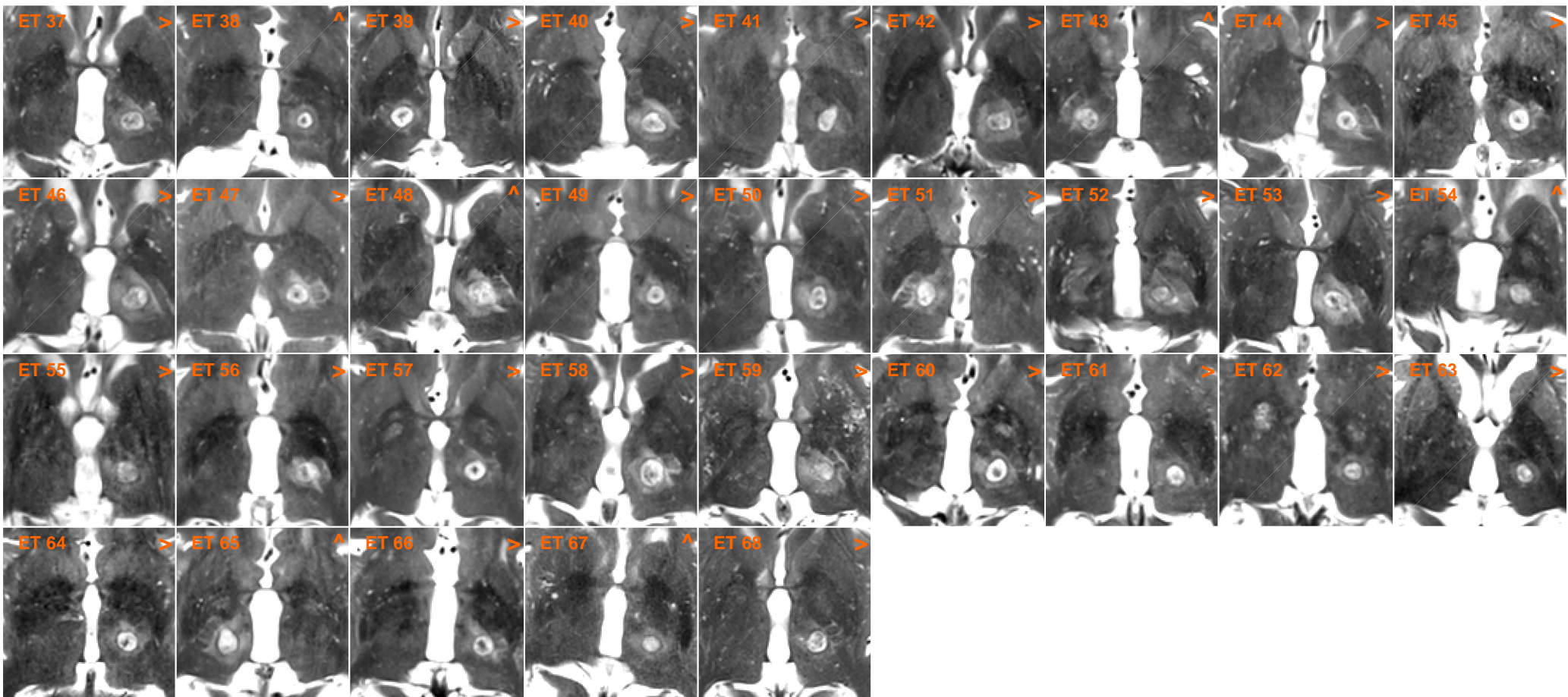

**ET Patients 1-36**
**Coronal T2WI; 24h after MRgFUS**

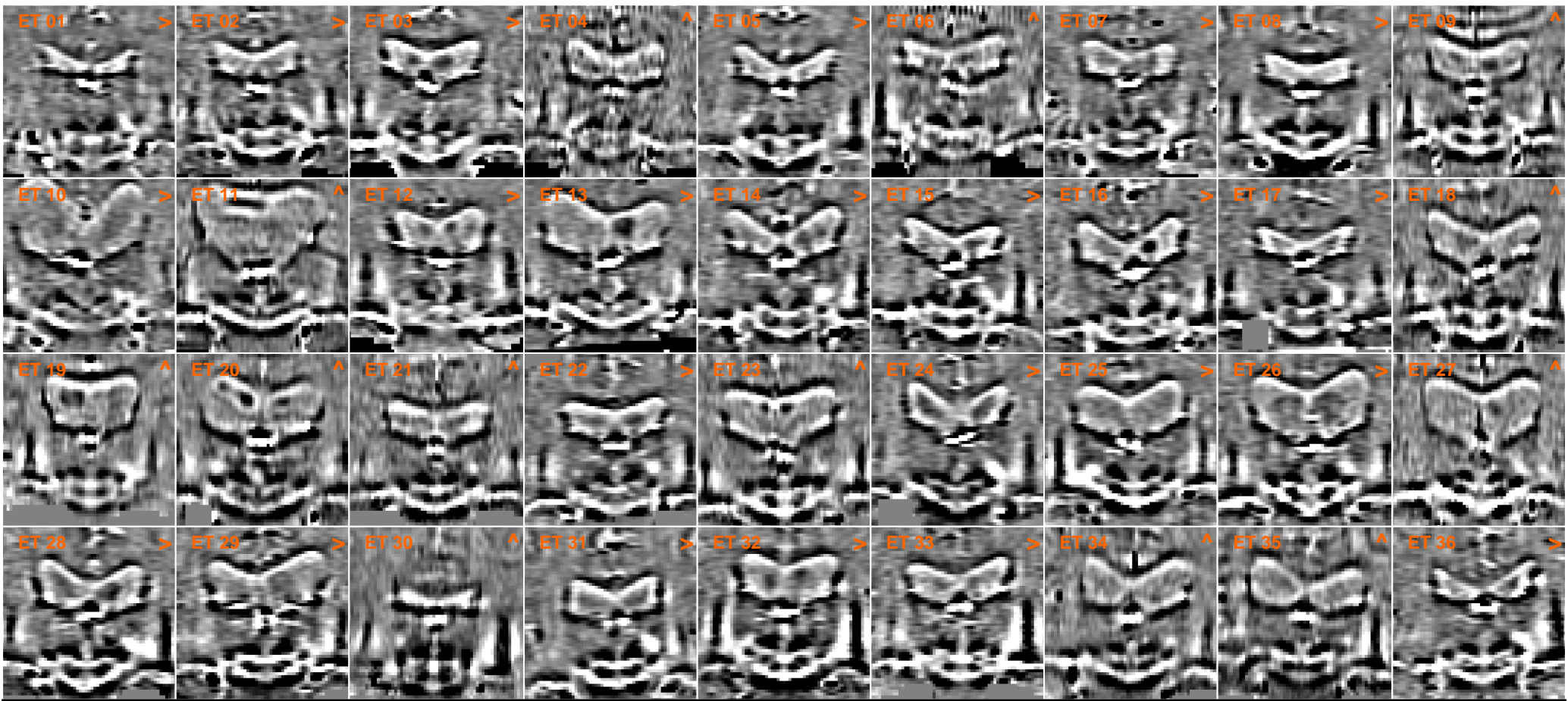

**ET Patients 1-36**
**Coronal T2WI; 24h after MRgFUS**

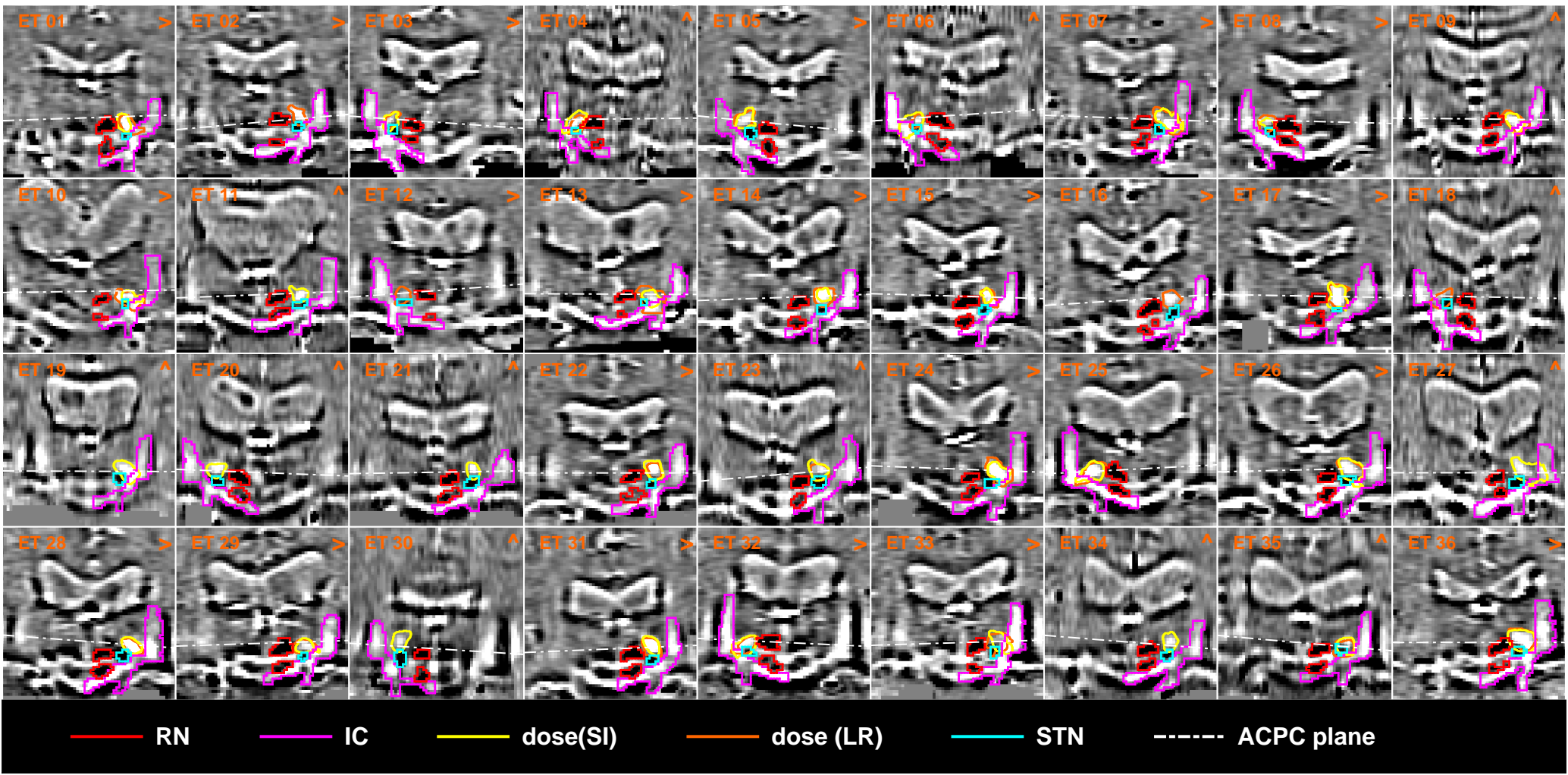

**ET Patients 1-36**
**Coronal T2WI; 24h after MRgFUS**

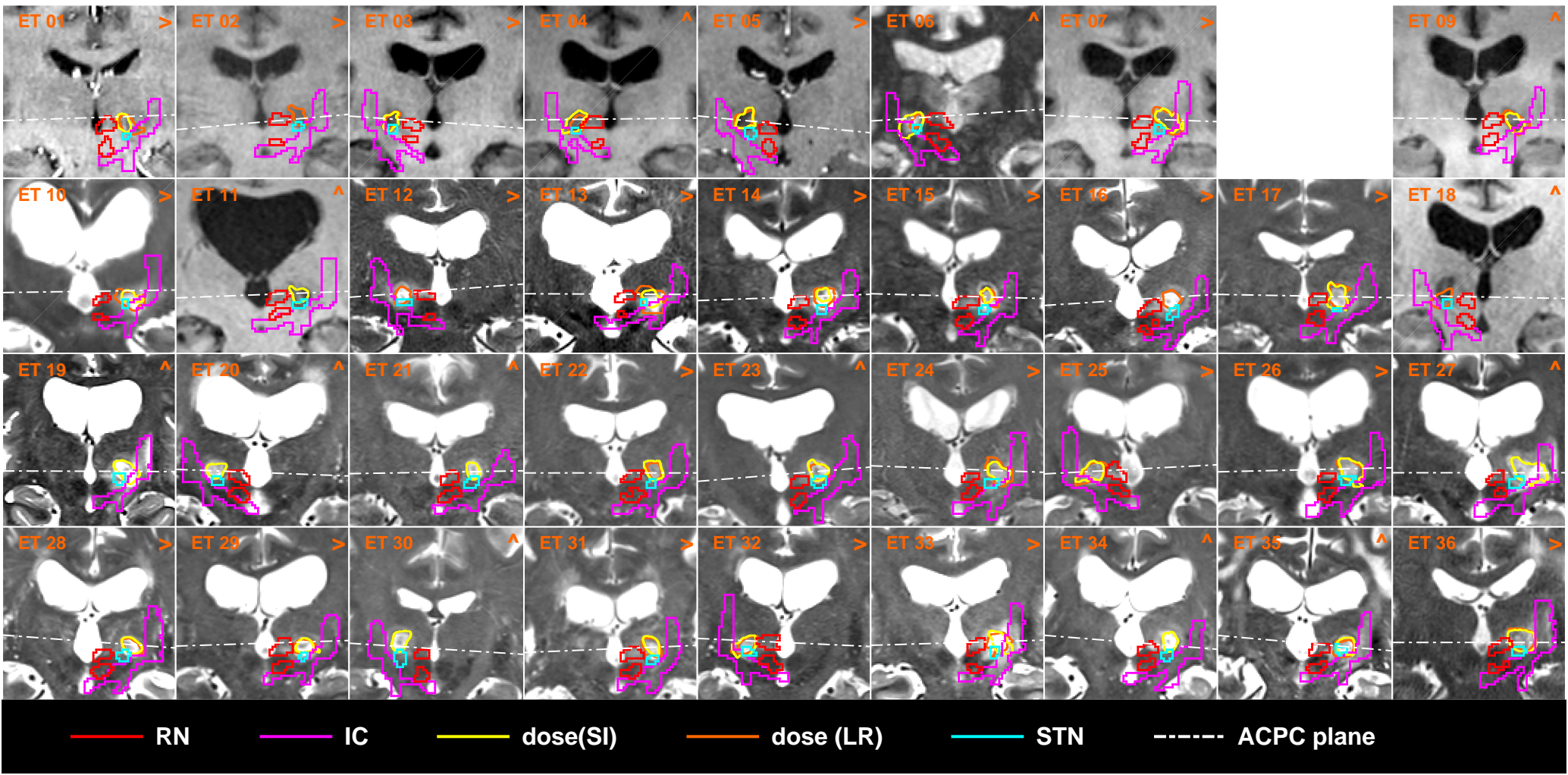

**ET Patients 1-36**
**Coronal T2WI; 24h after MRgFUS**

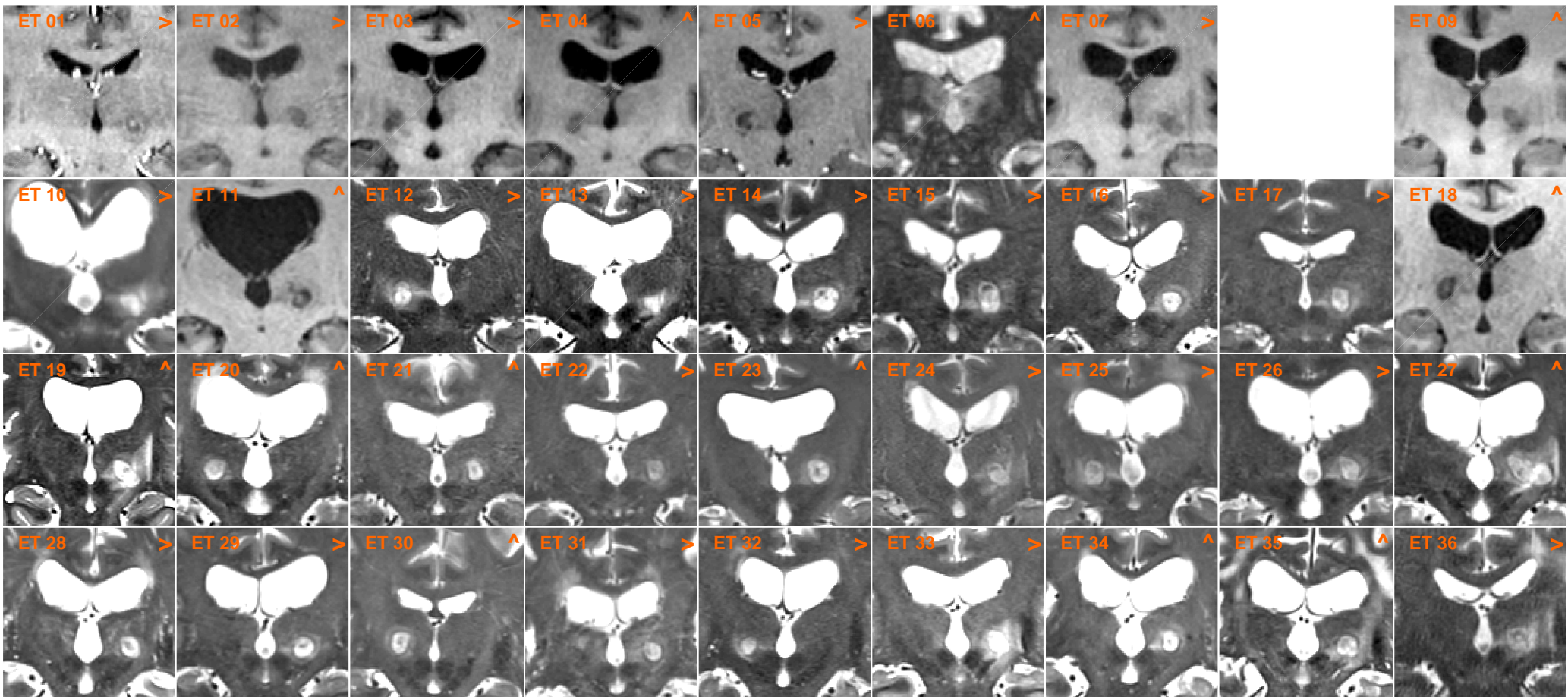

| RN | IC | dose(SI) | dose (LR) | STN | ACPC plane |

**ET Patients 37-68**
**Coronal T2WI; 24h after MRgFUS**

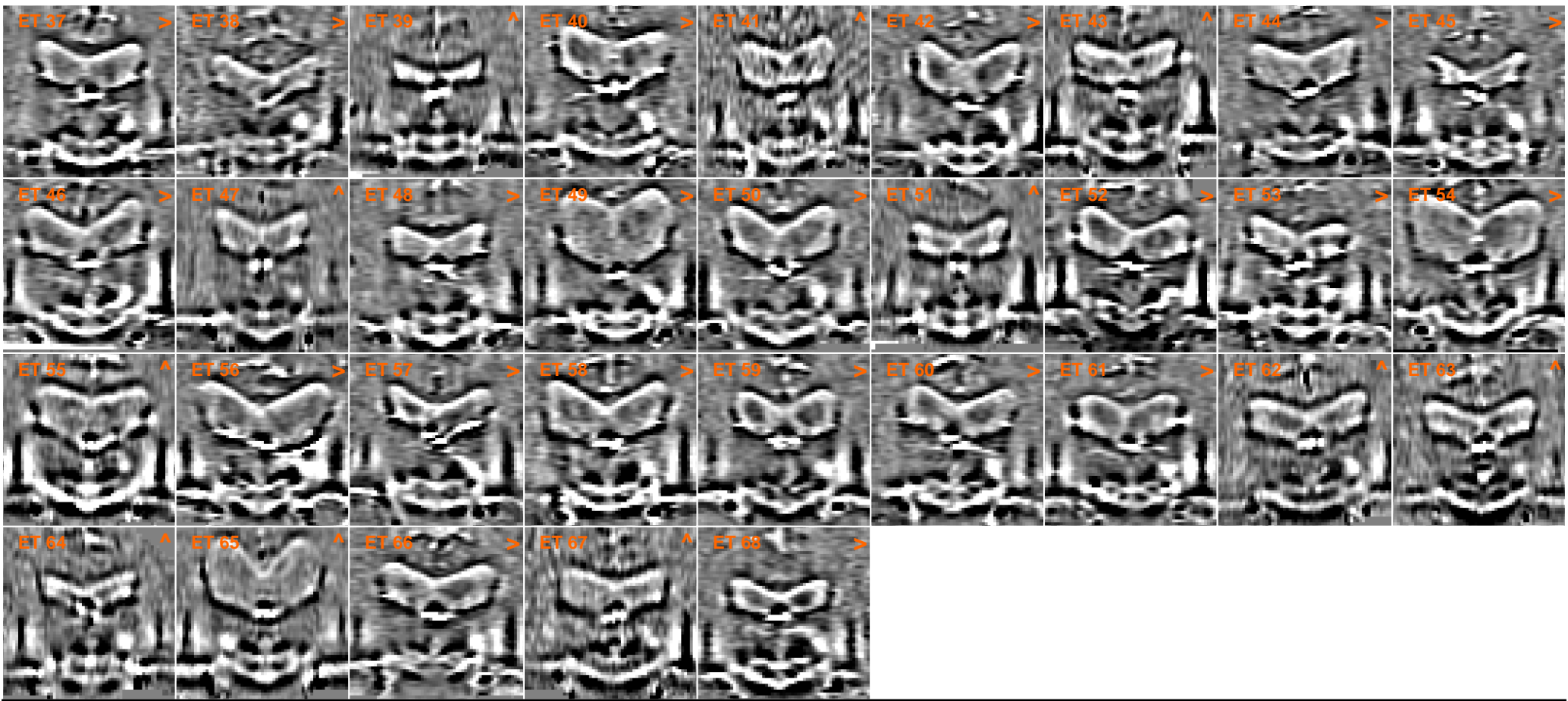

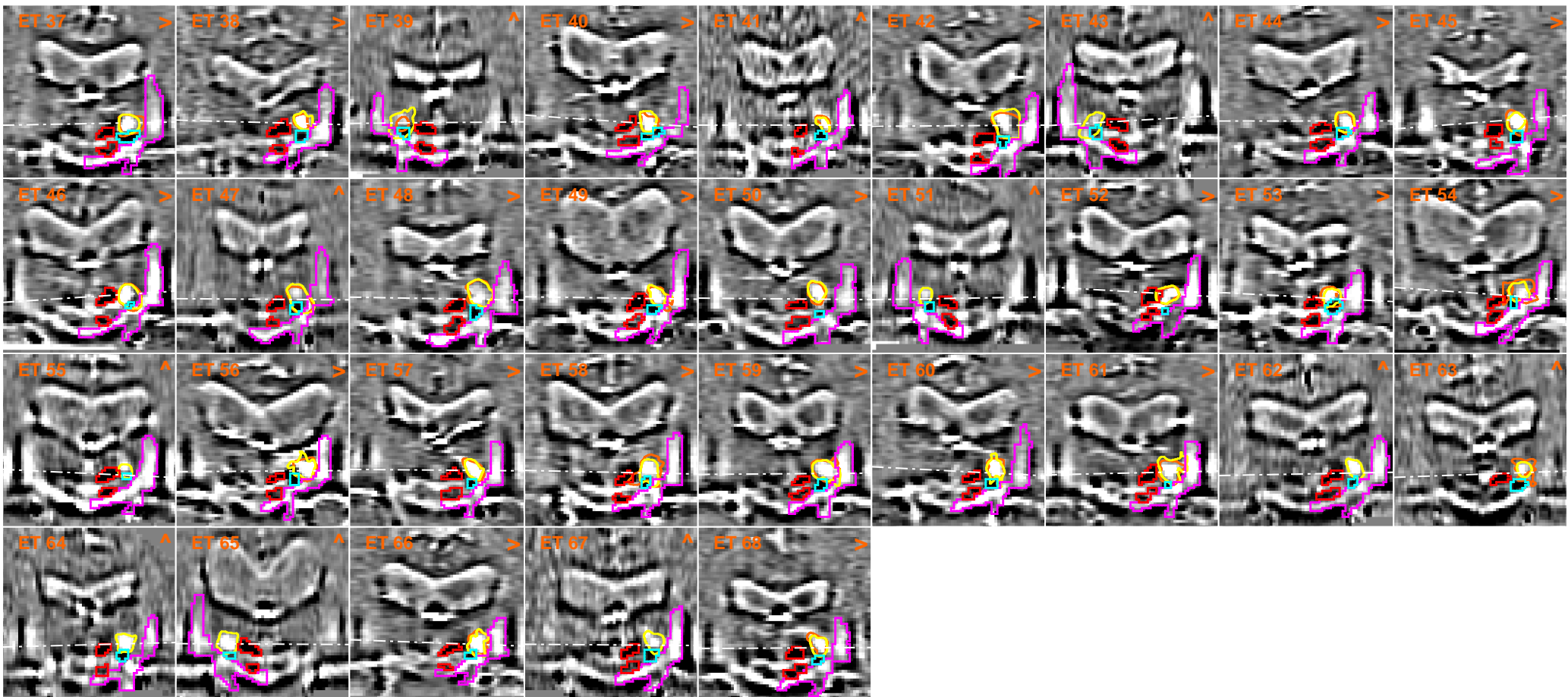

**ET Patients 37-68**
**Coronal T2WI; 24h after MRgFUS**

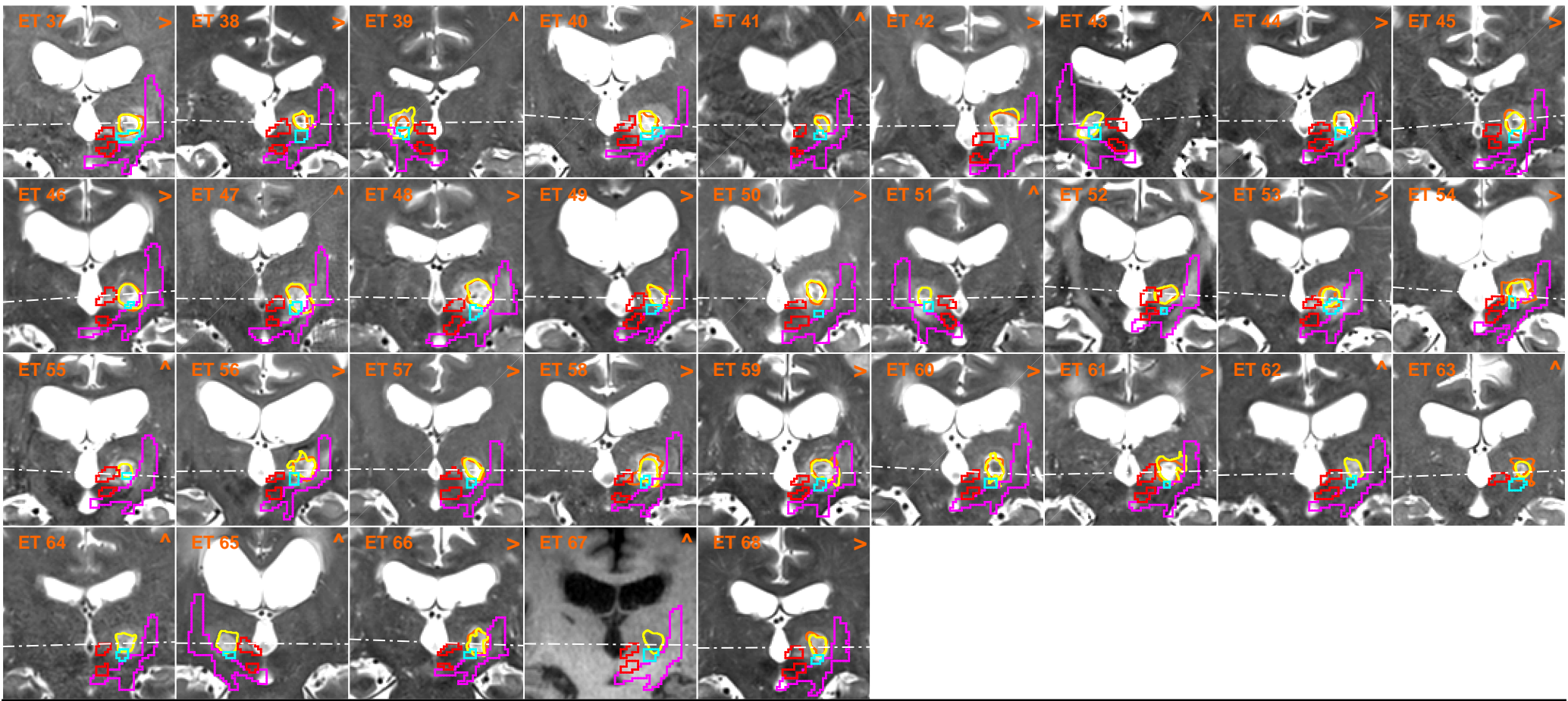

**ET Patients 37-68**
**Coronal T2WI; 24h after MRgFUS**

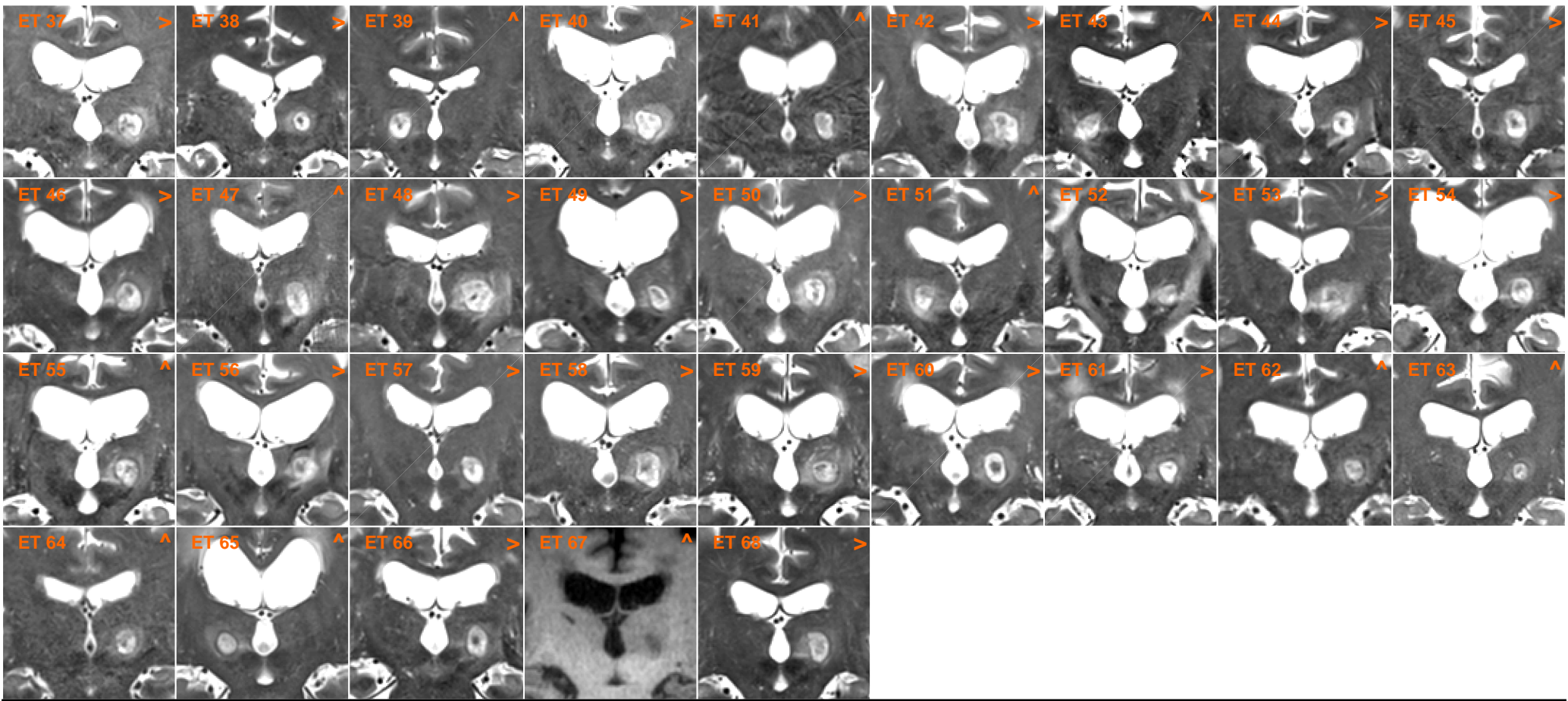

RN    IC    dose(SI)    dose (LR)    STN    ACPC plane